\documentclass[useAMS,usenatbib,usegraphicx]{mn2e}
\pdfoutput=1
\usepackage[UKenglish]{babel}
\usepackage[utf8]{inputenc}
\usepackage{amsmath,amsfonts,amssymb}
\usepackage{mathtools}
\DeclareSymbolFont{bbold}{U}{bbold}{m}{n}
\DeclareSymbolFontAlphabet{\mathbbold}{bbold}
\usepackage{natbib}
\usepackage{graphicx}
\usepackage{hyperref}
\usepackage{breakurl}
\usepackage[dvipsnames,hyperref]{xcolor}
\usepackage{xspace}
\graphicspath{{./}{./figures/}}

\definecolor{DiagDarkBlue}{RGB}{0,69,134}      
\definecolor{DiagOrange}{RGB}{255,66,14}       
\definecolor{DiagYellow}{RGB}{255,211,32}      
\definecolor{DiagGreen}{RGB}{87,157,28}        
\definecolor{DiagDarkViolet}{RGB}{126,0,33}    
\definecolor{DiagLightBlue}{RGB}{131,202,255}  
\definecolor{DiagDarkGreen}{RGB}{49,64,4}      
\definecolor{DiagLightGreen}{RGB}{174,207,0}   
\definecolor{DiagViolet}{RGB}{75,31,111}       
\definecolor{DiagGolden}{RGB}{255,149,14}      
\definecolor{DiagRed}{RGB}{197,0,11}           
\definecolor{DiagBlue}{RGB}{0,132,209}         

\colorlet{HighlightColor}{DiagDarkBlue}
\colorlet{MyLinkColor}{HighlightColor}
\colorlet{MyURLColor}{HighlightColor}
\colorlet{MyCiteColor}{HighlightColor}
\colorlet{MyFileColor}{HighlightColor}
\colorlet{MyMenuColor}{HighlightColor}
\colorlet{MyPageColor}{HighlightColor}

\hypersetup{%
	colorlinks=true,%
	bookmarksnumbered=true,%
	bookmarksopen=false,%
	breaklinks=true,%
	pdfborder={0 0 0},%
	pdftitle={Quality of restored images in OLBI},%
	pdfauthor={Nuno Gomes, Paulo J.\ V.\ Garcia, Éric Thiébaut},%
	unicode=true,%
	linkcolor=MyLinkColor,%
	urlcolor=MyURLColor,%
	citecolor=MyCiteColor,%
	filecolor=MyFileColor,%
	menucolor=MyMenuColor,%
}

\newcommand*{\Z}{\phantom{0}}
\newcommand*{\One}{\mathbbold{1}}
\newcommand*{\T}{^{\mathrm{T}}}
\DeclareMathOperator*{\argmin}{arg\,min}
\newcommand*{\softw}[1]{\textsc{\sf{#1}}}
\newcommand*{\MR}[1]{\ensuremath{\mathrm{#1}}}
\newcommand*{\Npix}{\ensuremath{N_\MR{pix}}}
\newcommand*{\acc}{\ensuremath{\text{ACC}}}
\newcommand*{\kl}{\ensuremath{\text{KL}}}
\newcommand*{\ibc}{\ensuremath{\text{IBC}}}
\newcommand*{\fid}{\ensuremath{\text{FID}}}
\newcommand*{\QualityIndex}{\ensuremath{\text{Q}}}
\newcommand*{\psnr}{\ensuremath{\text{PSNR}}}
\newcommand*{\mse}{\ensuremath{\text{MSE}}}
\newcommand*{\lonen}{\ensuremath{\text{L1N}}}
\newcommand*{\ltwon}{\ensuremath{\text{L2N}}}
\newcommand*{\wltwon}{\ensuremath{\text{WL2N}}}
\newcommand*{\ssim}{\ensuremath{\text{SSIM}}}
\newcommand*{\given}{\ensuremath{\:\vert\:}}
\newcommand*{\cf}{cf.\xspace}
\newcommand*{\arrs}[1]{\renewcommand{\arraystretch}{#1}} 
\newcommand{\mira}{\softw{MiRA}\xspace}
\newcommand{\newuoa}{\softw{NEWUOA}\xspace}

\makeatletter
\newcommand*{\etc}{\@ifnextchar{.}{etc}{etc.\@\xspace}}
\makeatother

\makeatletter
\newcommand*{\eg}{e.g.,\@\xspace}
\makeatother
\makeatletter
\newcommand*{\ie}{i.e.,\@\xspace}
\makeatother

\newcommand*{\Set}[1]{\mathbb{#1}}
\newcommand*{\Integers}{\Set{Z}}
\newcommand*{\Reals}{\Set{R}}

\newcommand*{\Extend}[1]{\widetilde{#1}}

\title[Assessing the quality of restored images]{Assessing the quality of restored images in optical long-baseline interferometry}
\author[Gomes, Garcia \& Thiébaut]{Nuno Gomes$^{1, 2, 3}$\thanks{E-mail:
nunogomes@fe.up.pt}, Paulo J.\ V.\ Garcia$^{1, 2}$ and Éric Thiébaut$^{4}$\\
$^{1}$ Universidade do Porto - Faculdade de Engenharia, Rua Dr. Roberto Frias, 4200-465 Porto, Portugal\\
$^{2}$ CENTRA, Instituto Superior Técnico, Av. Rovisco Pais, 1049-001 Lisboa, Portugal\\
$^{3}$ Universidade do Porto - Faculdade de Ciências, Rua do Campo Alegre, 4169-007 Porto, Portugal\\
$^{4}$ Centre de Recherche Astrophysique de Lyon/Obvservatoire de Lyon, 9 Avenue Charles André, 69561 Saint-Genis Laval Cédex, France}

\begin{document}

	\date{Accepted 2016 November 06. Received 2016 October 27; in original form 2016 July 19}
	\pagerange{\pageref{firstpage}--\pageref{lastpage}} \pubyear{2016}
	\maketitle
	\label{firstpage}

	\begin{abstract}
		Assessing the quality of aperture synthesis maps is relevant for benchmarking image reconstruction algorithms, for the scientific exploitation of data from optical long-baseline interferometers, and for the design/upgrade of new/existing interferometric imaging facilities.
		Although metrics have been proposed in these contexts, no systematic study has been conducted on the selection of a robust metric for quality assessment.
		This article addresses the question: what is the best metric to assess the quality of a reconstructed image?
		It starts by considering several metrics and selecting a few based on general properties.
		Then, a variety of image reconstruction cases are considered.
		The observational scenarios are phase closure and phase referencing at the Very Large Telescope Interferometer (VLTI), for a combination of two, three, four and six telescopes.
		End-to-end image reconstruction is accomplished with the \mira software, and several merit functions are put to test.
		It is found that convolution by an effective point spread function is required for proper image quality assessment.
		The effective angular resolution of the images is superior to naive expectation based on the maximum frequency sampled by the array.
		This is due to the prior information used in the aperture synthesis algorithm and to the nature of the objects considered.
		The $\ell_1$ norm is the most robust of all considered metrics, because being linear it is less sensitive to image smoothing by high regularization levels.
		For the cases considered, this metric allows the implementation of automatic quality assessment of reconstructed images, with a performance similar to human selection.
	\end{abstract}


	\section{Introduction}
	\label{sec:introduction}
	Existing optical long-baseline interferometers provide information at angular scales a factor of 10 smaller than any existing or planed single aperture telescope.
	This is achieved by measuring interference fringes from pairs of telescopes.
	The fringes' contrast and position at the detector can be related to the spatial coherence of the incoming electromagnetic field, which in turn contains information on the object brightness distribution (\cf \eg \citealt{Buscher2015,Glindemann2011}).
	This makes an imaging interferometer very different from an imaging camera.
	The first difference is related to the information content.
	A camera generates an image from a continuous sampled pupil, while an interferometer only obtains information at a much smaller number of specific locations of an effective `meta-pupil' -- the so-called $uv$-coverage of the data.
	A second difference is that while in a camera all the information is obtained simultaneously, in an interferometer data is taken from diverse array combinations separated in time.
	Finally, for an interferometer an algorithm must be used to synthesize an image.
	
	In optical long-baseline interferometry, phase information degradation by atmospheric turbulence is normally overcome by phase closure triangulation (\eg \citealt{Jennison1958,Monnier2007}), at the expense of further reducing the information content of the measurement.
	It is therefore not surprising that the first optical long-baseline images were of binaries (morphological simple objects) and were first obtained with three telescopes (\citealt{Baldwin1996,Benson1997}).
	Since the publication of the first relevant results, the technique of image reconstruction of long-baseline interferometric data in the optical/infrared (O/IR; 0.4--20\,\micron) regime has evolved and it is nowadays well established.
	A major breakthrough in optical long-baseline interferometry was the availability of the CHARA and Very Large Telescope Interferometer (VLTI) arrays \citep{tenBrummelaar2005,Schoeller2007} coupled to the control of atmospheric effects with spatial filtering \citep{CoudeduForesto1997,Tatulli2010} and adaptive optics \citep[\eg][]{Arsenautl2003}.
	By combining three or more telescopes and reasonable $uv$-coverages, the information content allowed us to overcome the binary barrier and enter into more complex morphologies such as stellar surfaces and discs (\eg \citealt{Benisty2011,Che2011,Hillen2016,Kloppenborg2015,LeBouquin2009,Millour2011,Mourard2015}).
	
	Because of the low information content of interferometric data, the generation of images is an ill-posed problem with more unknowns than available data.
	Therefore, images are reconstructed by minimizing a cost function that includes both the data and some prior information on the object brightness distribution (\eg \citealt{Thiebaut2013}).
	To overcome the effects of the turbulence, optical long-baseline interferometry data traditionally rely on the closure phase (and not on the baseline phase).
	The non-convex nature of the problem makes image reconstruction a difficult task, and algorithms are still a matter of active research (\cf \citealt{Berger2012} for a recent review).
	The availability of dispersed fringes increased the information content of interferometry data, enabling spectral self-calibration (\eg \citealt{Millour2011,Schutz2014}).
	Other developments are algorithms joining imaging and parametric descriptions of the astronomical objects (\eg \citealt{Kluska2014}), or different types of regularization \citep{Renard2011,Baron2014}.
	
	With the advent of \textit{GRAVITY} at European Southern Observatory, the first common instrument allowing phase referencing observations \citep{Eisenhauer2008}, most of the aperture synthesis algorithms may be simplified, because when a reference source is available, the phase closure is no longer required to remove atmospheric effects and the baseline phase becomes accessible.
	Standard radio interferometry approaches have proved successful with simulated data in this context (\eg \citealt{Vincent2011}).
	
	The large variety of aperture synthesis methods naturally leads to the question on which is the best approach.
	In 2001, the Working Group on Optical Interferometry of the International Astronomical Union (IAU) decided to compare and promote the development of different algorithms to restore O/IR interferometric images on a regular basis.
	Starting in 2004, an `Imaging Beauty Contest' has been held by SPIE every two years \citep{Lawson2004,Lawson2006,Cotton2008,Malbet2010,Baron2012,Monnier2014a}, where contestants present blindly restored images from synthetic or observational data provided by the organization of the contest.
	They are also asked to interpret the results, indicating what is believed to be real features and what are the potential artefacts of the imaging process.
	Subsequently, the restored images obtained from the different software are compared to their corresponding reference images by means of a best-fitting method.
	This method typically comprises a resampling of the restored image to the grid of the reference one, the normalization of the restored image to its peak brightness, and the comparison with the reference image convolved with the effective point spread function (PSF) of the interferometer, using a root-mean-square agreement.
	However, this approach is limited, because a particular metric might favour a special algorithm for a specific object morphology.
	This is a pertinent objection which, to our knowledge, is not addressed in the literature.
	
	The work presented here addresses this very question: how can we equitably measure the quality of an image obtained in aperture synthesis?
	This is a topic of relevance not only for algorithms, but also to the scientific exploitation of aperture synthesis, and for any future infrastructure relying on aperture synthesis imaging, such as the Planet Formation Imager (\citealt{Monnier2014b,Kraus2014}).
	
	This article is structured as follows.
	In Section~\ref{sec:image_quality}, we review merit functions used for image quality assessment, and we select a few for further analysis.
	It is underlined that image convolution with an effective PSF is mandatory.
	In Section~\ref{sec:methods}, we present the methods we used to recover the interferometric images, explaining how we generate the observables and respective noise, how we restored the images, and how we assess their quality.
	Important aspects of this approach are (\textit{a}) both phase closure and phase referencing techniques are addressed, and (\textit{b}) the array configurations are selected from available stations at the VLTI, particularly the case for four telescopes using phase closure, where the configurations are the ones used with the \textit{PIONIER} instrument.
	Section~\ref{sec:results} concerns about the reconstructed images and the analysis of the behaviour of the selected merit functions.
	We discuss the results and provide a summary of our findings.
	The most surprising outcome is that the metric used in the `Imaging Beauty Contest' is biased, but it can be replaced by a simple metric.
	A side bonus of our approach is that it paves the way for image quality assessment without human intervention.
	In Section~\ref{sec:conclusions} we conclude and present directions for future developments.

	\section{Image quality}
	\label{sec:image_quality}
	The quality of an image has to be assessed by an objective quantitative criterion.
	What is the best criterion also largely depends on the context.
	Here we will assume that the \emph{metric} $\Theta(x, y)$ is used to estimate the discrepancy between a reconstructed image $x$ and a reference image $y$.
	To simplify the discussion, we also assume that the lower the $\Theta(x, y)$ the better the agreement between $x$ and $y$.
	In other words, $\Theta(x, y)$ can be thought as a measure of the distance between $x$ and $y$.
	
	When assessing image quality, it is important that the result does not depend on irrelevant changes.
	This, however, depends on the type of images and on the context.
	For instance, for object detection or recognition, the image metric should be insensitive to the background level, to a geometrical transform (translation, rotation, magnification, \etc) or to a multiplication of the brightness by some positive factor which does not affect the shape of the object.
	In cases where image reconstruction has underdeterminations, these should not have any incidence on the metric.
	For optical interferometry and when only power-spectrum and closure phase data are available, the images to be compared may have to be shifted for best matching.
	In general, the metric should be minimized with respect to the undetermined parameters.
	
	When comparing a true image $z$ (with potentially an infinitely high resolution) to a restored image $x$, the effective resolution achievable by the instrument and the image restoration process must be taken into account.
	Otherwise and because image metrics are in general based on pixel-wise comparisons, the slightest displacement of sharp features would lead to large loss of quality (according to the metric) whereas the images may look very similar at a lower and more realistic resolution.
	The easiest solution is then to define the reference image $y$ to be the true image $z$ blurred by an effective PSF $h_\MR{ref}$, whose shape corresponds to the effective resolution
	\begin{equation}
	\label{eq:reference-image}
		y = h_\MR{ref} \ast z ,
	\end{equation}
	where the symbol asterisk $(\ast)$ denotes the convolution.
	The choice of the effective resolution is then a parameter of the metric.
	
	To summarize and to be specific, using the distance $\Theta(x, y)$ between the restored image $x$ and the reference image $y$, the discrepancy between $x$ and the true image $z$ would be given by:
	\begin{equation}
	\label{eq:general-discrepancy}
		d(x, z) = \min_{\alpha,\beta,\sigma,t}
			\Theta\bigl(\alpha\,h_{\sigma,t} \ast x + \beta, h_\MR{ref} \ast z \bigr),
	\end{equation}
	with $\alpha$ a brightness scale, $\beta$ a background, and $h_{\sigma,t}$ a matching PSF of width parameter\footnote{In this paper we took $\sigma$ to be the standard deviation of the PSF profile.} $\sigma > 0$ and centred at position $t$.
	Note that the merit function should be minimized with respect to the width $\sigma$ of the effective PSF in order to estimate the effective resolution achieved by a given restored image.
	Our choice to assigning the translation to the matching PSF is to avoid relying on some particular method to perform sub-pixel interpolation (of $x$, $y$ or $z$) for fine tuning the position.
	Not doing so would add another ingredient to the metric.
	When dealing with images with different pixel sizes, the resampling of the images at a given common resolution can be implemented by a linear operator which performs at the same time the resampling, the fine shifting and the blurring by one of the PSFs.
	
	In the following subsections, we first review the most common metrics found in the literature and argue whether they are appropriate or not in the context of optical interferometry.
	We then propose a family of suitable metrics.

		\subsection{Merit functions}
		\label{sec:merit_functions}
			
			\subsubsection{Quadratic metrics}
			\label{sec:quadratic_metric}
			Quadratic merit functions are probably the most widely used ones, for they are easy to manipulate and can be made insensitive to various effects, such as an affine change in the image levels (see Section~\ref{sec:affine-correction}).
			Even though it is not always obvious, they are, in fact, related to various metrics proposed for comparing images.
			Compared to the Kullback--Leibler divergence (see Section~\ref{sec:kl}), quadratic merit functions amount to assuming a simple distribution of the differences between two images (that is to say, independent and Gaussian).
			The most general expression of a quadratic metric to measure the discrepancy between two images $x$ and $y$ takes the form of a weighted (squared) $\ell_2$-norm:
			\begin{displaymath}
				\wltwon(x, y; W) = \Vert x - y \Vert_{W}^2 ,
			\end{displaymath}
			where we denote by $\Vert q \Vert_{W}^2 = q\T \, W \, q$ the weighted squared Euclidean norm, with $W$ a positive (semi-)definite weighting operator.
			Using a diagonal weighting operator $W = \MR{diag}(w)$ yields:
			\begin{equation}
			\label{eq:quad}
				\wltwon(x, y; w) = \sum_{i} w_i \, (x_i - y_i)^2,
			\end{equation}
			where the sum is carried out for all pixels of the images and where the $w_i \ge 0$ is the weight of pixel $i$.
			
			By choosing specific weights, it is possible to mimic a number of commonly used metrics.
			For instance, the metric of the \textit{Interferometric Imaging Beauty Contest} \citep{Lawson2004} is
			\begin{align}
			\label{eq:ibc}
				\ibc(x, y)
					&= \sqrt{\wltwon(x, y; w = y/{\textstyle \sum_{i} y_{i}})} \nonumber \\
					&= \left[\frac{\sum_{i} y_i \, (x_i - y_i)^2}{\sum_{i} y_i}\right]^{1/2} ,
			\end{align}
			which amounts to taking the weights as being proportional to the reference image: $w = y/\sum_{i} y_{i}$.
			The main drawbacks of this merit function are that it overemphasizes the brighter regions of the image and discards pixels where the reference image $y$ is zero, which occurs for many pixels for a compact astronomical source on a dark background.
			For these reasons, we anticipate that $\ibc$ may not be the best metric.
			
			The most simple quadratic metric is the squared $\ell_2$-norm (also known as the \textit{squared Euclidean norm}) of the pixel-wise differences between the images:
			\begin{align}
			\label{eq:l2-norm}
				\ltwon (x, y)
					&= \Vert x - y \Vert_2^2 \nonumber \\
					&= \sum_{i} (x_i - y_i)^2 ,
			\end{align}
			which is $\wltwon$ when $w = 1$.
			The \textit{Mean Squared Error} (MSE) is directly derived from the Euclidean norm by taking $w = 1/\Npix$, with $\Npix$ the number of pixels:
			\begin{equation}
			\label{eq:mse}
				\mse(x, y) = \frac{1}{\Npix} \, \lVert x - y \rVert_2^2 .
			\end{equation}
			The MSE was used by \citet{Renard2011} to benchmark the effects of the regularization in the image reconstruction from interferometric data.
			For all the metrics presented so far, the smaller the merit value, the more similar are the images.
			
			Some other commonly used metrics are also based on the Euclidean norm of the differences.
			For instance, the \textit{Peak Signal to Noise Ratio} (PSNR) is
			\begin{equation}
			\label{eq:psnr}
				\psnr(x, y) = 10 \times \log_{10} \! \left(
					\frac{ \bigl[ \max(y) - \min(y) \bigr]^2 }{ \mse(x, y) }
				\right) .
			\end{equation}
			Here, $\min(y)$ and $\max(y)$ correspond respectively to the minimum and maximum possible pixel value of the reference image $y$.
			The PSNR is given in decibel (db) units and the higher the PSNR, the more similar are the images.
			
			Clearly, MSE and PSNR are the squared Euclidean norm of the pixel-wise difference between the images ($\ltwon$) but expressed in different units.
			They can be used interchangeably and we will only consider $\ibc$ and $\ltwon$ in what follows.

			\subsubsection{Minimizing the discrepancy with respect to the brightness distortion}
			\label{sec:affine-correction}
			In order to make a formal link between different metrics, it is worth investigating what happens when the minimization with respect to the brightness distortion parameters $\alpha$ and $\beta$ is carried on.
			As we will show, this minimization has a closed form solution with a quadratic metric:
			\begin{displaymath}
				\Vert \alpha \, x + \beta\,\One - y \Vert_{W}^2 ,
			\end{displaymath}
			with $x$ and $y$ the images to compare, $\alpha \in \Reals^+$ a positive factor, $\beta \in \Reals$ a constant background, and $\One$ an image where all pixels are equal to 1.
			
			Let us first consider the constant background correction.
			Introducing $r = y - \alpha \, x$, we want to minimize $\Vert r - \beta\,\One\Vert_{W}^2$ with respect to $\beta$.
			Expanding the quadratic norm yields
			\begin{displaymath}
				\Vert r - \beta\One\Vert_{W}^2 = \Vert r \Vert_{W}^2
				- 2\,(\One\T\,W\,r)\,\beta + \Vert \One \Vert_{W}^2\,\beta^2 .
			\end{displaymath}
			This is a simple 2$^\MR{nd}$ order polynomial in $\beta$ and the minimum is achieved for the optimal background
			\begin{equation}
			\label{eq:optimal_background}
				\beta^\star = \frac{\One\T\,W\,r}{\One\T\,W\,\One} ,
			\end{equation}
			which can be seen as a weighted averaging of $r$.
			Thus,
			\begin{equation}
				\min_{\beta} \Vert r - \beta\,\One\Vert_{W}^2
					= \Vert r - \beta^\star\,\One\Vert_{W}^2
					= \Vert C\,r\Vert_{W}^2 ,
			\end{equation}
			where the linear operator $C$ is given by
			\begin{equation}
			\label{eq:centering-operator}
				C = I - \One\,\frac{\One\T\,W}{\One\T\,W\,\One} ,
			\end{equation}
			and $I$ is the identity.
			The linear operator $C$ has the effect of removing the weighted average of its argument.
			Replacing $r$ by $y - \alpha \, x$ yields:
			\begin{equation}
				\min_{\beta} \Vert  \alpha \, x + \beta\,\One - y \Vert_{W}^2
					= \Vert \alpha\,C\,x - C\,y\Vert_{W}^2 ,
			\end{equation}
			which amounts to comparing the weighted average subtracted images.
			
			The expansion
			\begin{displaymath}
				\Vert \alpha \, x - y \Vert_{W}^2
					= \Vert y \Vert_{W}^2 - 2\,(y\T\,W\,x)\,\alpha + \Vert x \Vert_{W}^2 \, \alpha^2
			\end{displaymath}
			readily shows that the optimal factor $\alpha$ is
			\begin{displaymath}
				\argmin_{\alpha} \Vert \alpha \, x - y \Vert_{W}^2
					= \frac{y\T\,W\,x}{x\T\,W\,x} ,
			\end{displaymath}
			and, after trivial simplifications, that
			\begin{displaymath}
				\min_{\alpha} \Vert \alpha \, x - y \Vert_{W}^2
					=  \Vert y \Vert_{W}^2 - \frac{(y\T\,W\,x)^2}{\Vert x \Vert_{W}^2} .
			\end{displaymath}

			Putting all together we have shown that
			\begin{equation}
				\min_{\alpha,\beta} \Vert \alpha \, x + \beta\,\One - y \Vert_{W}^2
					= \Vert C\,y \Vert_{W}^2 - \frac{(y\T\,C\T\,W\,C\,x)^2}{\Vert C\,x \Vert_{W}^2} ,
			\end{equation}
			where the linear operator $C$ is given in equation~(\ref{eq:centering-operator}).
			If no background correction is wanted, it is sufficient to take $C = I$.
			The above expression can be divided by $\Vert C\,y \Vert_{W}^2$ to obtain a symmetric distance between $x$ and $y$ which is independent of an affine transform of the brightness of any of the two images
			\begin{equation}
				d(x, y) = 1 - \MR{Corr}(x, y)^2 \, ,
			\end{equation}
			with
			\begin{equation}
			\label{eq:weighted-correlation}
				\MR{Corr}(x, y) = \frac{y\T\,C\T\,W\,C\,x}
					{\Vert C\,x \Vert_{W} \, \Vert C\,y \Vert_{W}}
			\end{equation}
			the (weighted) correlation between the two images $x$ and $y$.
			If $W \propto I$, then the usual definition of the correlation, given in equation~(\ref{eq:correlation}), is retrieved.
			
			The distance $d(x, y)$ takes values in the range $[0,1]$, the smaller it is the better is the agreement.
			Conversely, the better the agreement the larger the absolute value of the (weighted) correlation.
			It is therefore clear now that comparing images by means of their (weighted) correlation coefficient is equivalent to using a quadratic norm minimized with respect to an affine transform of the image intensity.

			\subsubsection{Universal image quality index and image structural similarity}
			\label{sec:quality_index}
			\label{sec:ssim}
			The \textit{universal image quality index} was proposed by \citet{Wang2002} to overcome MSE and PSNR, which were found to be very poor estimators of the image quality for common brightness distortions and image corruptions (like salt-and-pepper noise, lossy compression artefacts, \etc).
			The universal image quality index is defined as
			\begin{equation}
			\label{eq:quality_index}
				\QualityIndex(x, y) = \frac{
					4\,\MR{Avg}(x)\,\MR{Avg}(y)\,\MR{Cov}(x,y)
				}{
					\bigl(\MR{Avg}(x)^2 + \MR{Avg}(y)^2\bigr)\,
					\bigl(\MR{Var}(x) + \MR{Var}(y)\bigr)
				} ,
			\end{equation}
			where $\MR{Avg}(x)$, $\MR{Var}(x)$ and $\MR{Cov}(x,y)$ are respectively the empirical average, variance and covariance of $x$ and $y$, given by:
			\begin{align*}
				\MR{Avg}(x) &= \frac{1}{\Npix} \, \sum_{i} x_i \, , \\
				\MR{Var}(x) &= \MR{Cov}(x,x) \, , \\
				\MR{Cov}(x, y) &= \frac{1}{\Npix - 1} \, \sum_{i}
				\bigl(x_i - \MR{Avg}(x)\bigr) \, \bigl(y_i - \MR{Avg}(y)\bigr) \, .
			\end{align*}

			The universal image quality index takes values in the range $[-1,1]$.
			$Q(x,y)$ is maximal for the best agreement, which occurs when $y = \alpha\,x + \beta$, and minimal when $y = -\alpha\,x + \beta$, for any $\alpha > 0$ and any $\beta$.
			Although the universal image quality index was designed to cope with brightness distortions such as mean shift or dynamic shrinkage, this indicator is not exactly insensitive to any affine transform of the intensity as is (see the demonstration in Section~\ref{sec:affine-correction}) the correlation coefficient:
			\begin{equation}
			\label{eq:correlation}
				\MR{Corr}(x, y) = \frac{\MR{Cov}(x,y)}{\sqrt{\MR{Var}(x) \, \MR{Var}(y)}} .
			\end{equation}

			In order to improve over the universal image quality index,
			\citet{Wang2004} introduced the \textit{image Structural SIMilarity} (SSIM):
			\begin{align}
			\label{eq:ssim}
				\ssim(x, y) &=
				\frac{%
					2 \, \MR{Avg}(x) \, \MR{Avg}(y) + \varepsilon_1 %
				}{%
					\MR{Avg}(x)^2 + \MR{Avg}(y)^2 + \varepsilon_1
				} \notag \\
				& \quad \times \frac{
					2 \, \MR{Cov}(x, y) + \varepsilon_2%
				}{
					\MR{Var}(x) + \MR{Var}(y) + \varepsilon_2%
				} ,
			\end{align}
			where $\varepsilon_1 > 0$ and $\varepsilon_2 > 0$ are small values introduced to avoid divisions by zero.
			Note that with $\varepsilon_1 = 0$ and $\varepsilon_2 = 0$, the SSIM is just the image quality index defined in equation~(\ref{eq:quality_index}).
			The higher the SSIM, the better the agreement.
			In principle SSIM and the quality index should be used \emph{locally}, that is on small regions of the images.

			\subsubsection{Accuracy function}
			\label{sec:accuracy}
			Similarly to the IBC metric, the \textit{accuracy function} \citep[ACC, ][]{Gomes2016b} is based on a normalized weighted quadratic difference between the reconstructed image $x$ and the reference image $y$:
			\begin{equation}
			\label{eq:accuracy}
				\MR{ACC}(x, y) = \frac{
					\sum_{i} w_i \, (x_i - y_i)^2
					}{
					\sum_{i}  (x_i + y_i)^2
				} .
			\end{equation}
			Here $w$ is a normalized weighting function, a mask that eliminates all pixels where the reference and the restored images have intensities smaller than the image's dynamic range.
			On all non-negligible pixels, $w$ is equal to 1.
			
			ACC varies between 0 and 1 and the smaller it is, the greater the resemblance between both images.
			Note that the accuracy function is neither quadratic in $x$ nor in $y$.

			\subsubsection{Sum of absolute differences}
			\label{sec:l1-norm}
			One of the drawbacks of quadratic metrics is that they strongly emphasize the largest differences.
			To avoid this, an $\ell_p$-norm can be used with an exponent $p < 2$.
			For instance, the \textit{sum of absolute differences} or $\ell_1$-norm is given by:
			\begin{align}
			\label{eq:l1-norm}
				\lonen (x, y)
					&= \Vert x - y \Vert_1 \nonumber \\
					&= \sum_{i} \lvert x_i - y_i \rvert.
			\end{align}

			\subsubsection{Fidelity function}
			\label{sec:fidelity}
			The \textit{fidelity function} was introduced by \citet{Pety2001b} in the context of image reconstruction for ALMA.
			It is defined as the ratio of the total flux of the reference $y$ to the difference between the restored image $x$ and the reference one:
			\begin{equation}
			\label{eq:fidelity}
				\fid(x, y) = \frac{
					\sum_{i} y_i
				}{
					\sum_{i} \max \{ \eta, \vert y_i - x_i \vert \}
				} ,
			\end{equation}
			where $\eta$ is some non-negative threshold.
			The higher the fidelity value, the better the agreement.
			
			Choosing $\eta > 0$ avoids divisions by zero, and \citet{Pety2001b} took $\eta = 0.7 \, \MR{RMS}(x - y)$, where $\MR{RMS}(...)$ yields the root mean squared value of its argument.
			We note that with $\eta > 0$, all differences smaller than $\eta$ have the same incidence on the total cost and are therefore irrelevant.
			To avoid this, one has to take $\eta = 0$, in which case the reciprocal of the fidelity function is then just the $\ell_1$-norm defined in equation~(\ref{eq:l1-norm}) times some constant factor which only depends on the reference $y$.
			As the fidelity function would then yield the same results as the $\ell_1$-norm, we only consider the latter in our study.

			\subsubsection{Kullback--Leibler divergence}
			\label{sec:kl}
			Being non-negative everywhere and normalized, the images can be thought as
			distributions (over the pixels).
			The \textit{Kullback--Leibler divergence} measures the similarity between two distributions.
			When applied to our (normalized), images it writes
			\begin{equation}
			\label{eq:kl}
				\kl(x, y) = \sum_{i} y_i \, \log(x_i / y_i) .
			\end{equation}

			A restriction for the Kullback--Leibler divergence is that $x$ and $y$ must be strictly positive everywhere.
			It is however possible to account for non-negative distributions by modifying the definition of the Kullback--Leibler divergence as follows:
			\begin{displaymath}
				\kl(x, y) = \sum_{i} c_{\kl}(x_i, y_i) ,
			\end{displaymath}
			where $c_{\kl}(q, r)$ extends $r \log(q/r)$ by continuity:
			\begin{displaymath}
				c_{\kl}(q, r) = \
				\begin{cases}
					0             & \text{if $q = r$, or $q > 0$ and $r = 0$,}\\
					-\infty       & \text{if $q = 0$ and $r > 0$,} \\
					r \log(q / r) & \text{otherwise.}
				\end{cases}
			\end{displaymath}
			Note that the Kullback--Leibler divergence is not symmetric, \ie $\kl(x,y) \not= \kl(y,x)$.
			The Kullback--Leibler divergence is less or equal to zero.
			The lower the Kullback--Leibler divergence the worse is the agreement between $x$ and $y$.
			The maximal value of the Kullback--Leibler divergence is equal to zero and is achieved when $x = y$.
			
			Like the IBC metric, the Kullback--Leibler divergence disregards $x_i$ where $y_i = 0$.
			In addition, any image $x$ with at least one pixel, say $i_0$, such that $x_{i_0} = 0$ while $y_{i_0} > 0$ yields $\kl(x, y) = -\infty$, which corresponds to the maximum possible discrepancy.
			These are serious drawbacks for using the Kullback--Leibler divergence as an image metric, because it could not make a distinction between restored images such that $x_{i_0} = 0$, whatever the values of the other pixels.

			\subsubsection{Designing the metric}
			\label{sec:designing_metric}
			We want to derive an image metric that is adapted to our particular case: we consider images of compact objects (\ie with finite size support) over a constant background, and which may be shifted by an arbitrary translation.
			
			We assume that $d(x, y, t)$ yields the discrepancy between the image $x$ and the image $y$ shifted by a translation $t$.
			Quite naturally, we require that the following properties hold:
			\begin{itemize}
				\item[(\textit{i})] The metric does not change if the images are extended with pixels set with the background level;
				likewise, the metric does not change if the images are truncated, provided that the values of the removed pixels equal the background level;
				\item[(\textit{ii})] The metric is non-negative and equal to zero if the two images are the same (for a given relative translation);
				in particular $d(x, x, 0)= 0$, whatever the image $x$;
				\item[(\textit{iii})] The metric is \emph{stationary} in the sense that whatever the images $x$ and $y$ and the translations $t$, $t'$ and $t''$,
				\begin{equation}
					d\bigl(s(x,t),s(y,t'),t''\bigr) = d(x,y,t+t''-t') ,
				\end{equation}
				where $s(x,t)$ yields image $x$ shifted by translation $t$:
				\begin{displaymath}
					s(x,t)_{i} = x_{i - t} .
				\end{displaymath}
			\end{itemize}
			A last requirement, although optional, could be:
			\begin{itemize}
				\item[(\textit{iv})] The metric is \textit{symmetric} in the sense that
				\begin{equation}
				\label{eq:symmetry}
					d(y,x,-t) = d(x,y,t) ,
				\end{equation}
				whatever the images $x$ and $y$ and the translation $t$.
			\end{itemize}

			To limit the number of possibilities, we consider that the metric is the
			sum of a pixel-wise cost.
			Then, accounting for property~(\textit{i}),
			\begin{equation}
			\label{eq:additive_metric}
				d(x, y, t) = \sum_{i \in \Integers^n}
					c(\Extend{x}_{i}, \Extend{y}_{i - t}) ,
			\end{equation}
			where $n$ is the number of dimensions of the images $x$ and $y$ (in our case, $n= 2$), $\Integers$ is the set of integers, $t \in \Integers^n$ is the considered translation, $c(q, r)$ is the pixel-wise cost, and $\Extend{x}$ (resp.\ $\Extend{y}$) is the image $x$ (resp.\ $y$) infinitely extended with the background level $\beta$:
			\begin{equation}
				\Extend{x}_{i} =
					\begin{cases}
						x_{i} & \text{if $i \in \Set{X}$;} \\
						\beta & \text{else,} \\
					\end{cases}
			\end{equation}
			with $\Set{X} \subset \Integers^n$ (resp.\ $\Set{Y} \subset \Integers^n$) the support of the image $x$ (resp.\ $y$).
			We note that property~(\textit{ii}) implies that $c(q, q) = 0$ whatever $q \in \Reals$, and also that the background level must be the same for the two images.
			We also note that property~(\textit{iv}) implies that the pixel-wise cost be a symmetric function, \ie $c(q, r) = c(r, q)$ whatever $(q, r) \in \Reals^2$.
			Finally, property~(\textit{iii}) holds because the same pixel-wise cost is used whatever the index $i$.
			
			As $c(\beta,\beta) = 0$, the sum over the infinite set $\Integers^n$ in equation~(\ref{eq:additive_metric}) simplifies to sums over three finite (and possibly empty) subsets:
			\begin{equation}
			\label{eq:additive_metric_finite}
				d(x,y,t)
					= \sum_{\mathclap{i \in \Set{X} \cap \Set{Y}_{t}}} c(x_{i}, y_{i - t})
					+ \sum_{\mathclap{i \in \Set{X} \backslash \Set{Y}_{t}}} c(x_{i}, \beta)
					+ \sum_{\mathclap{i \in \Set{Y} \backslash \Set{X}_{t}}} c(y_{i}, \beta) ,
			\end{equation}
			where $\Set{A}\backslash\Set{B}$ denotes the set of elements of $\Set{A}$ which do not belong to $\Set{B}$, and
			\begin{displaymath}
				\Set{X}_{t} = \{ i \in \Integers^n \given i - t \in \Set{X} \}
			\end{displaymath}
			is the set of indices $i$ such that $i - t$ belongs to the support of $x$.
			An efficient implementation of the metric may be achieved with:
			\begin{equation}
			\label{eq:additive_metric_fast}
				d(x, y, t) = \gamma + \sum_{\mathclap{i \in \Set{X} \cap \Set{Y}_{t}}}
					\bigl[c(x_{i}, y_{i - t}) - c(x_{i}, \beta) - c(y_{i - t}, \beta)\bigr] ,
			\end{equation}
			where $c(x_{i}, \beta)$ (resp.\ $c(y_{i}, \beta)$) can be pre-computed for all $i\in\Set{X}$ (resp.\ for all $i\in\Set{Y}$) and
			\begin{displaymath}
				\gamma
					= \sum_{i \in \Set{X}} c(x_{i}, \beta)
					+ \sum_{i \in \Set{Y}} c(y_{i}, \beta) \, .
			\end{displaymath}

			Finally, it remains to choose the pixel-wise cost $c(q,r)$.
			A whole family of merit functions can be derived with the following pixel-wise cost
			\begin{equation}
				c(q,r) = \bigl\vert \Gamma(q) - \Gamma(r)\bigr\vert^p
			\end{equation}
			where $p > 0$ is a chosen exponent and $\Gamma$ is a function used to emphasize the discrepancy in the low/high range of the brightness distribution.
			For example, taking
			\begin{equation}
				\Gamma(q) = \mathrm{sign}(q)\,|q|^\gamma ,
			\end{equation}
			with $\gamma \in [0,1]$, it amounts to paying more attention to the least bright part of the images.
			Taking $p= 2$ and $\gamma= 1$ yields the $\ell_2$-norm (\ltwon), while taking the quadratic merit $p= 1$ and $\gamma= 1$ yields the $\ell_1$-norm (\lonen).
			Incidently, this shows that the required aforementioned properties (including the symmetry) do hold for these norms.

			\subsubsection{Choice of the candidates}
			\label{sec:choice_candidates}
			We already mentioned that not all merit functions reviewed in this paper are appropriate for comparing synthetic aperture images.
			For example, we disregarded the Kullback--Leibler divergence (see Section~\ref{sec:kl}) because of its inability to distinguish between very different images which have pixels equal to zero while they are non-zero in the reference image.
			In our context, the background level is known (\ie $\beta = 0$ which corresponds to the positivity constraint) and should not have to be adjusted when comparing images.
			The Universal Quality Index and Image Structural Similarity described in Section~\ref{sec:ssim} are therefore not appropriate for our needs.
            However, these metrics can be of value in image patches with non-zero backgrounds.\footnote{Using these metrics would also imply the definition of a patch size, which would open other questions outside the scope of this article.}
			The brightness scale $\alpha$ may have to be tuned so as to minimize the discrepancy between the images because, on the one hand, they may have different normalization constraints and, on the other hand, they may have been interpolated to cope with different pixel sizes.
			As we have shown in Section~\ref{sec:affine-correction}, minimizing a quadratic cost function in $\alpha$ would be equivalent to use the correlation of the images as a metric.
			
			To summarize, we will compare images using the $\ell_2$-norm (\ltwon), the $\ell_1$-norm (\lonen), the metric used in the past Interferometric Beauty Contests (\ibc) and the accuracy function (\acc).

	\section{Methods}
	\label{sec:methods}
	
		\subsection{Synthetic image library}
		\label{sec:image_library}
\begin{figure*}
			\begin{center}
				\includegraphics[height=5cm]{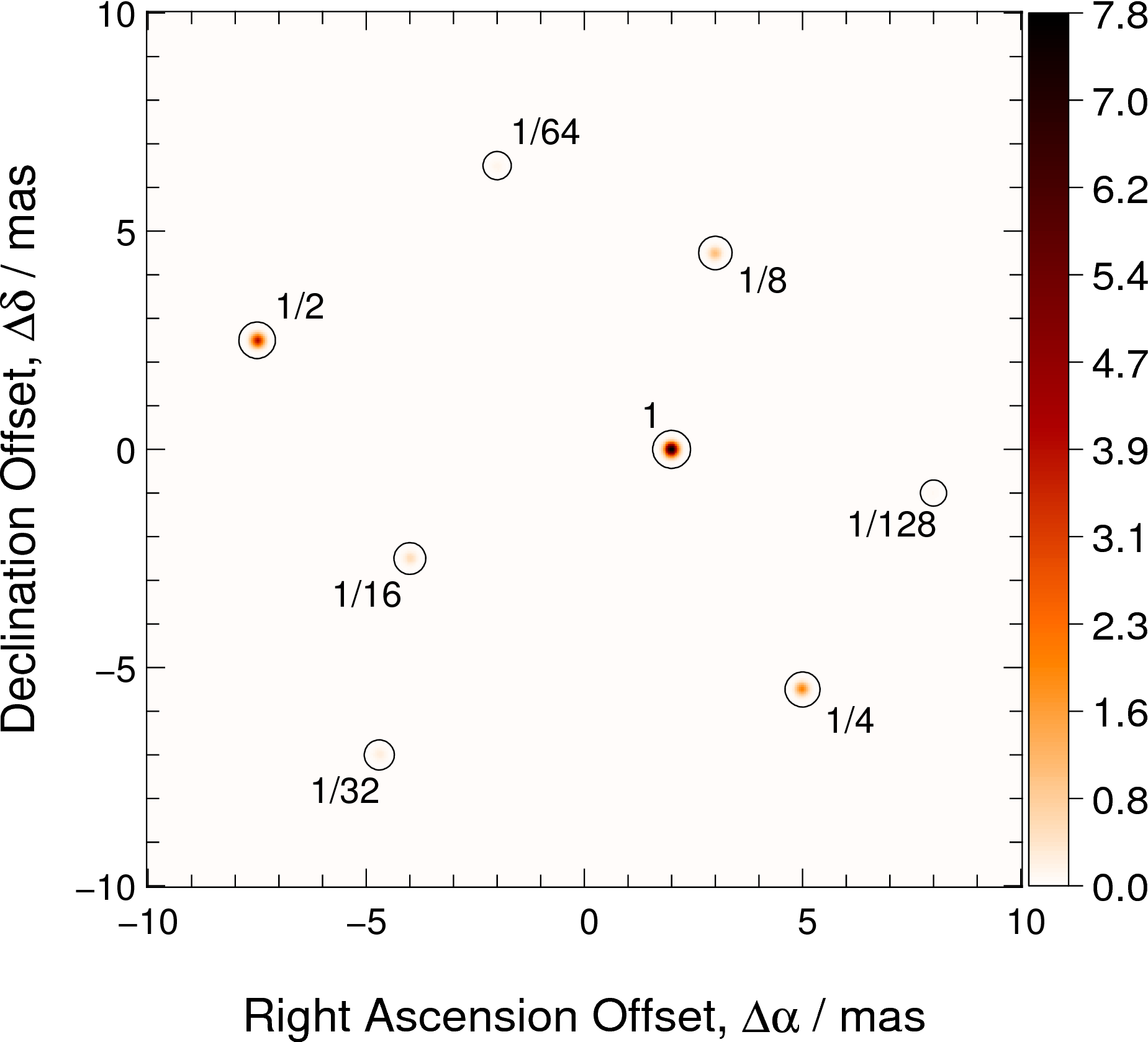}~
				\includegraphics[height=5cm]{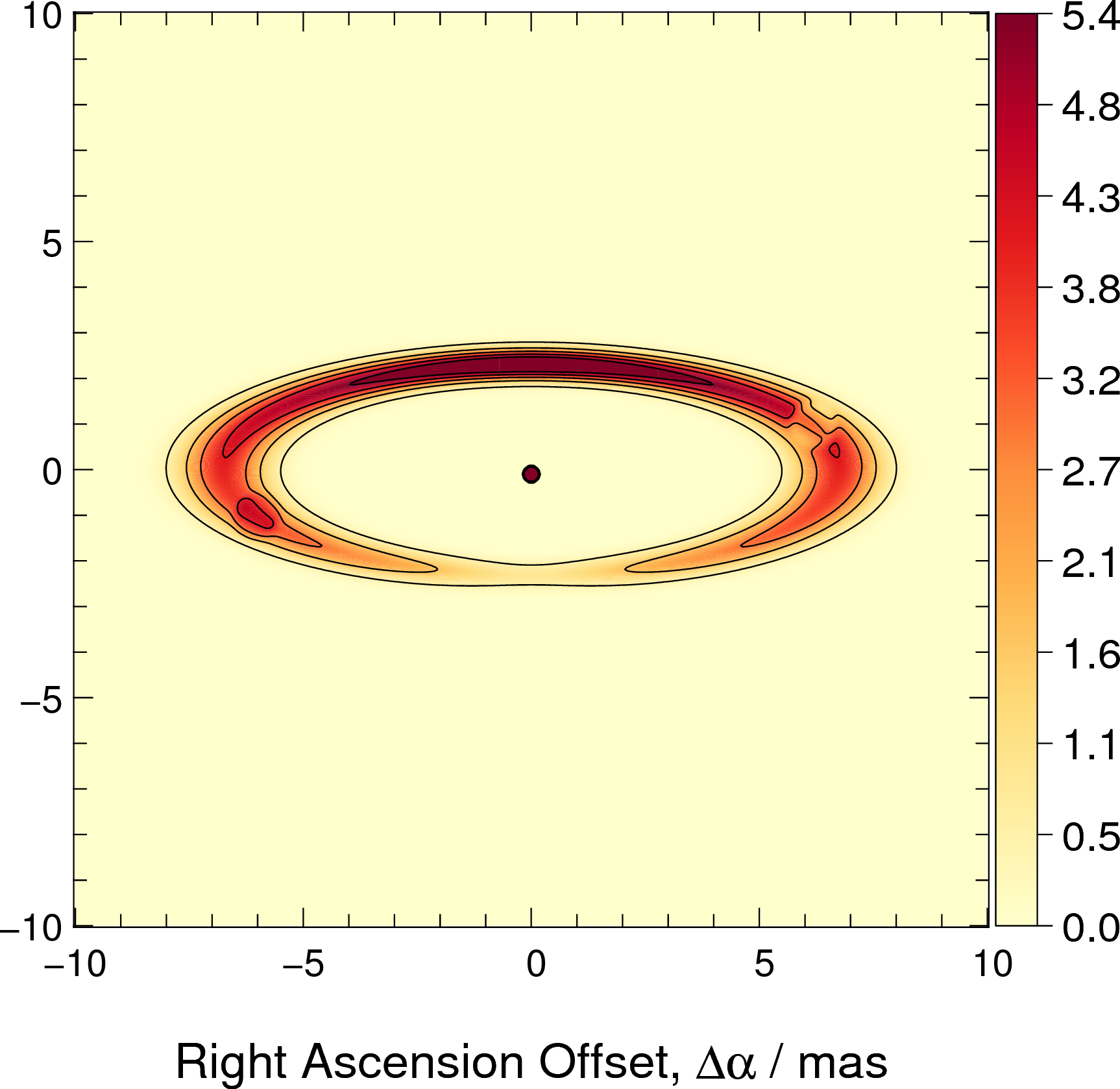}~
				\includegraphics[height=5cm]{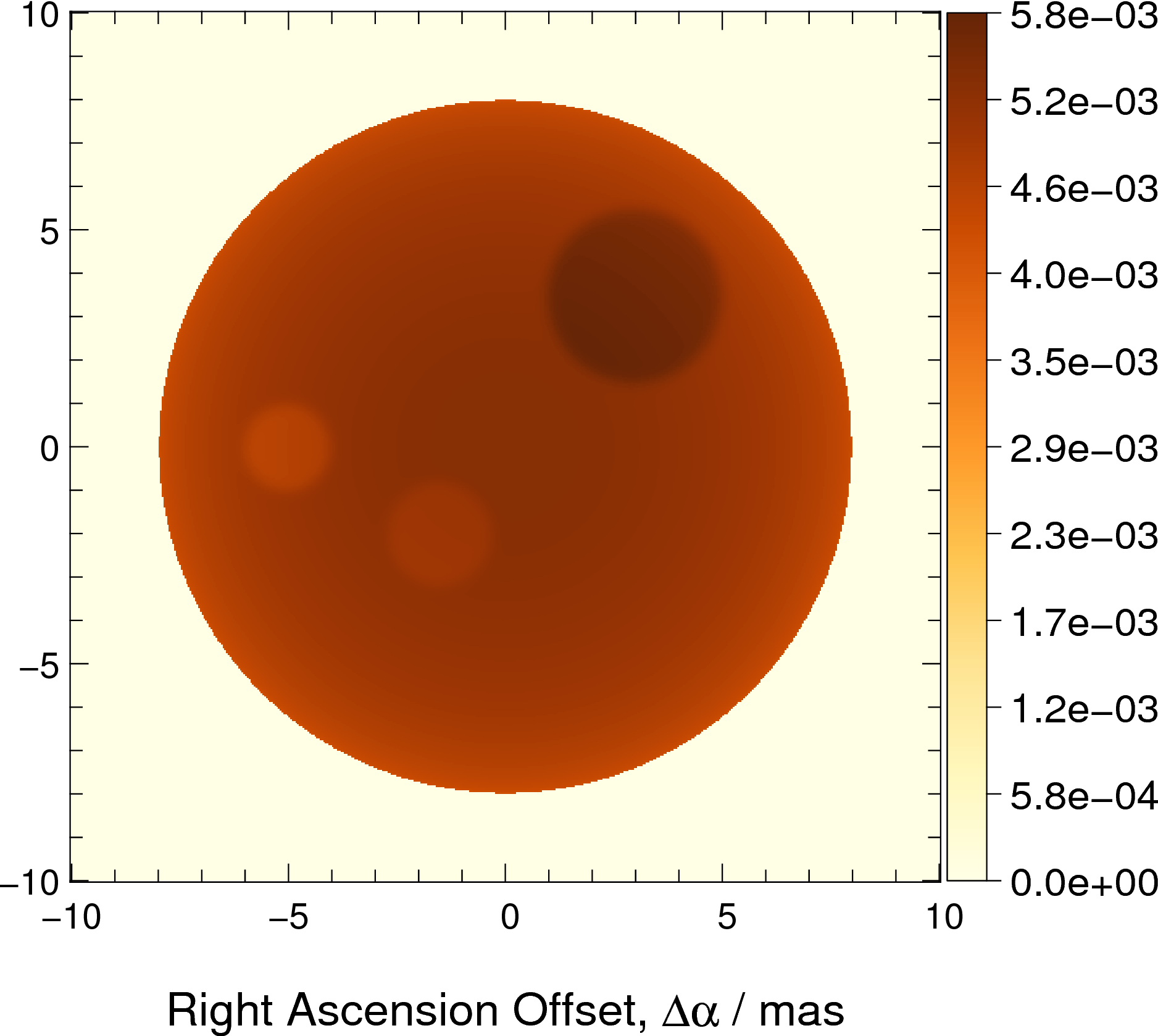}
				\caption{%
					\label{fig:reference}
					True images ($z$) used for the image reconstruction study: stellar cluster (\textit{left}), YSO \textit{centre}), and stellar photosphere (\textit{right}). %
					The images are normalized by their total flux. %
					The colour bars indicate surface flux. %
					The stars of the cluster have relative intensities as indicated in the figure. %
					The circles point the position of the stars. %
					The colour maps have been chosen in order to maximize the contrast of the features in each image.%
				}
			\end{center}
		\end{figure*}

		The true images ($z$) used in the study are presented in Fig.~\ref{fig:reference}.
		They span representative science cases of interferometric imaging (\cf \eg \citealt{Berger2012}): compact clusters/multiple stellar systems, young stellar objects (YSOs) and stellar surfaces.
		We fixed the size of the images to ease the interpretation of the results.
		The width of the pixel is 0.04\,mas.
		The images cover a wide range of visibilities, from the very sharp cluster to the over-resolved stellar photosphere.
		The cluster consists of eight stars `randomly' spread in the FOV, with a Gaussian profile of standard deviation 0.1\,mas, whose intensities decrease in factors of 2.
		The typical separation between neighbouring stars is 5\,mas.
		The YSO consists of a central star and a circumstellar disc, with a total flux ratio of 10 to 1.
		The disc has two features: a dark spot on the first quadrant and a bright spot in the third quadrant.
		The stellar surface has two bright spots in the third quadrant, and a dark spot on the first quadrant.

		\subsection{\texorpdfstring{$UV$}{uv}-space generation}
		\label{sec:uv-space_generation}
\begin{figure*}
			\begin{center}
			\includegraphics[height=5.5cm]{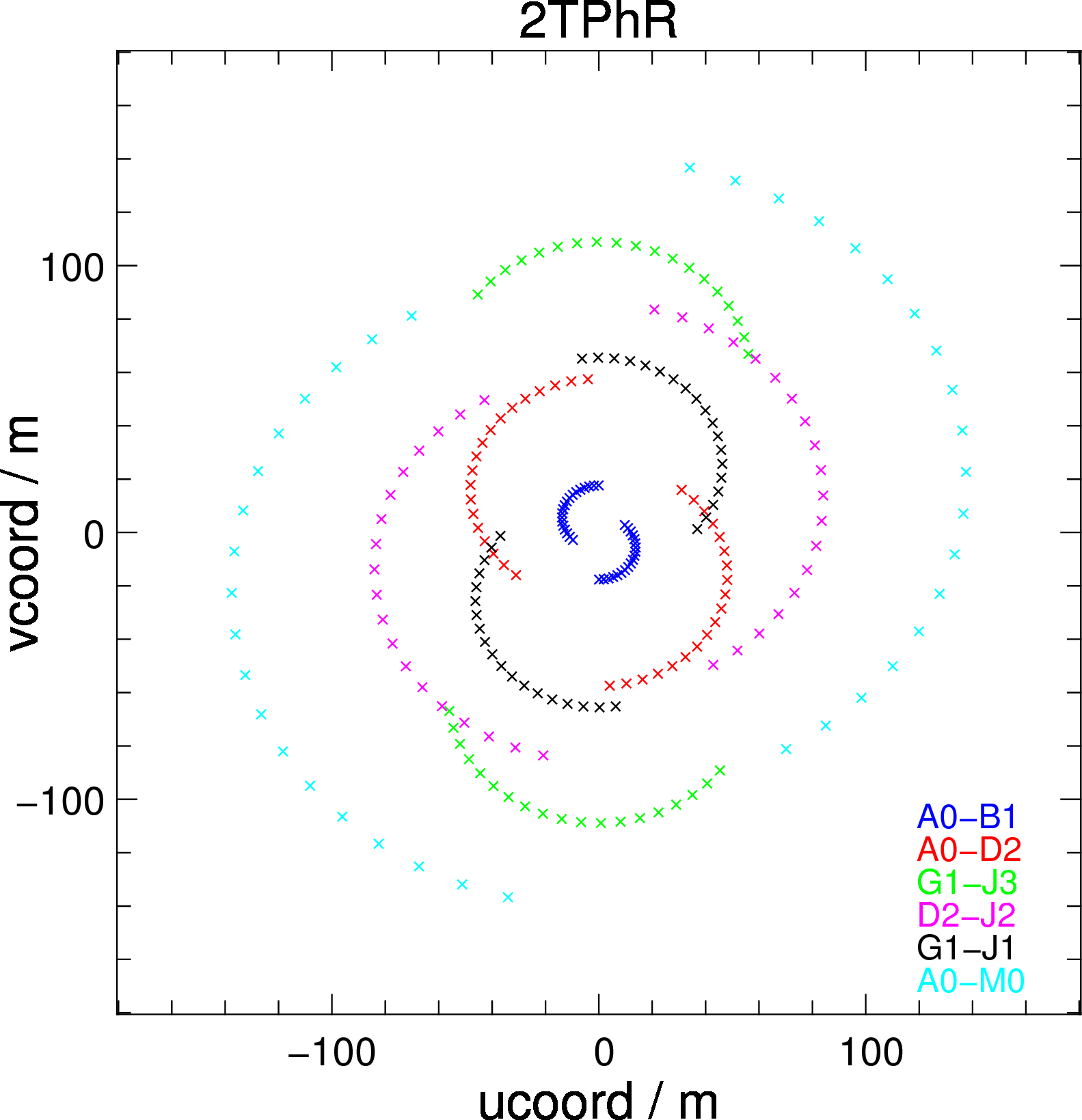}~
			\includegraphics[height=5.5cm]{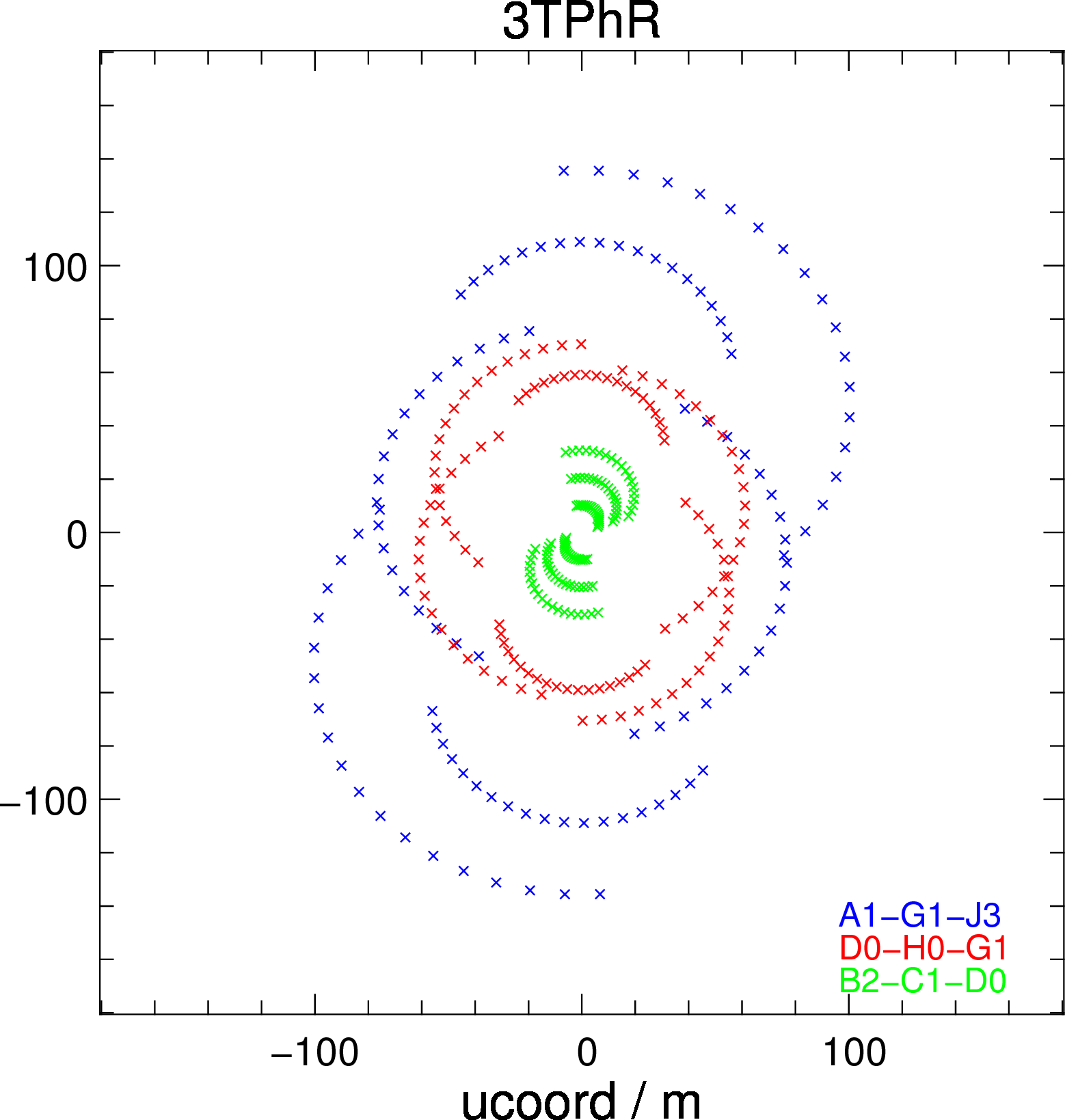}~
			\includegraphics[height=5.5cm]{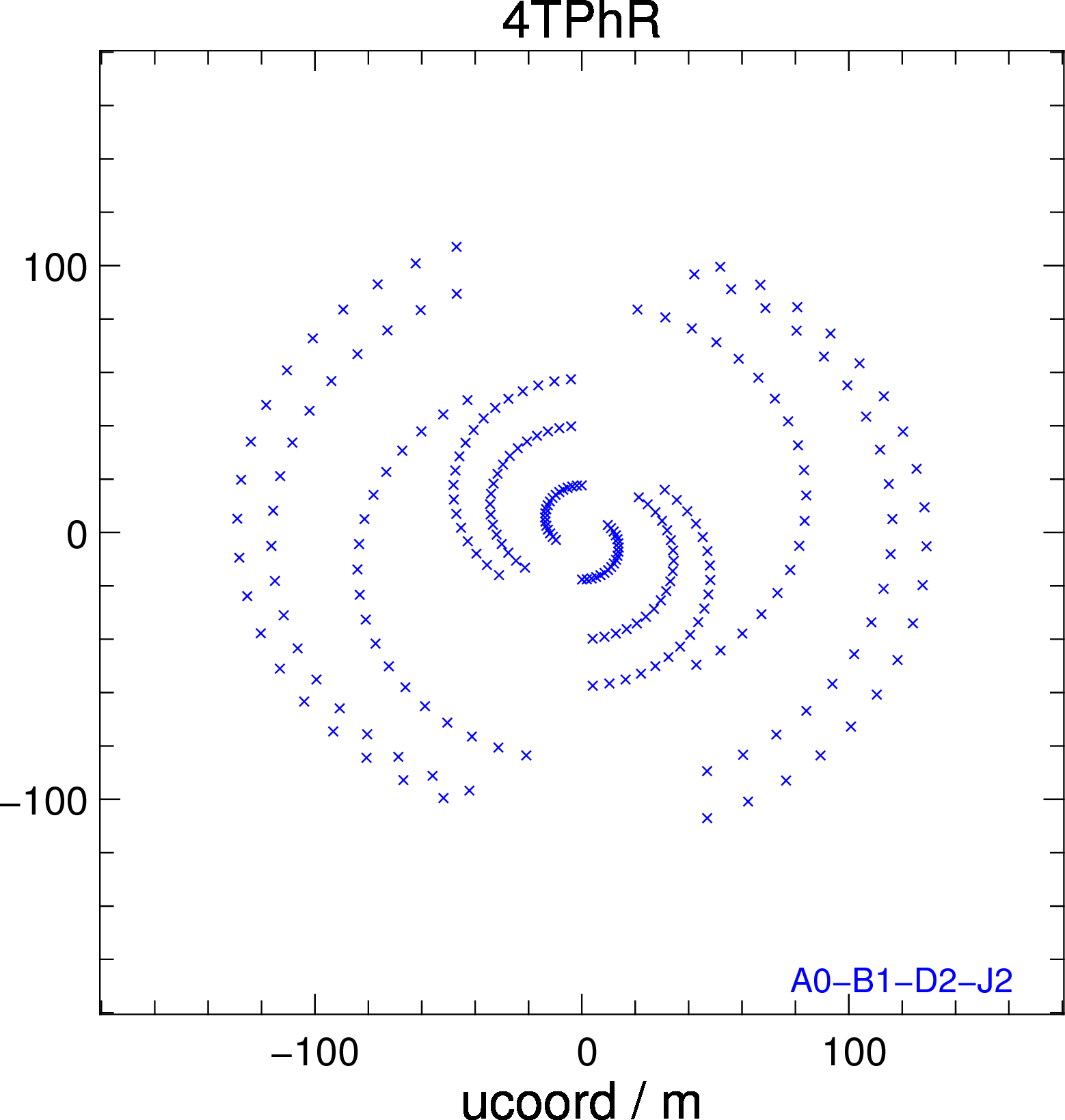}\\[8pt]
			\includegraphics[height=5.5cm]{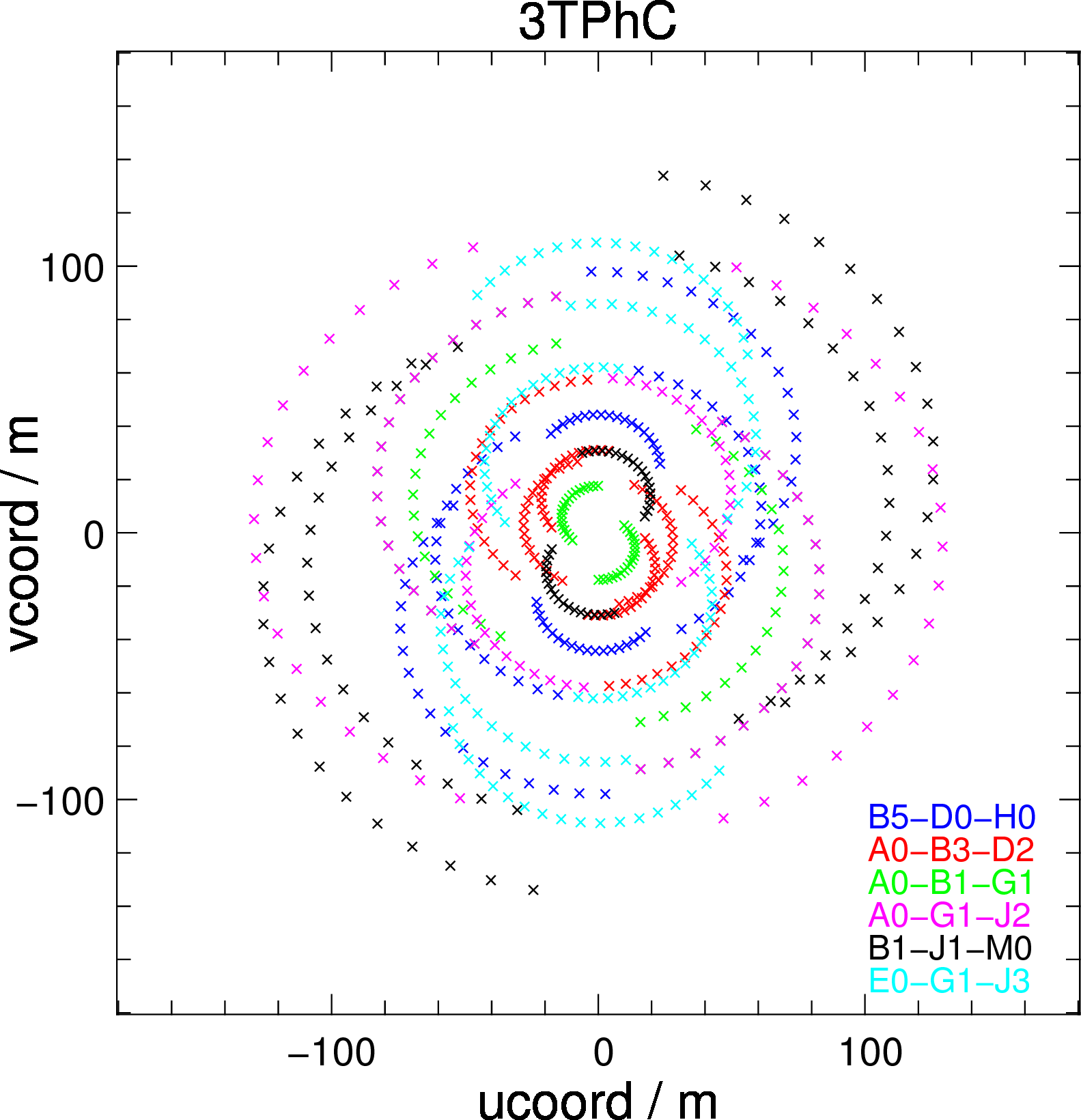}~
			\includegraphics[height=5.5cm]{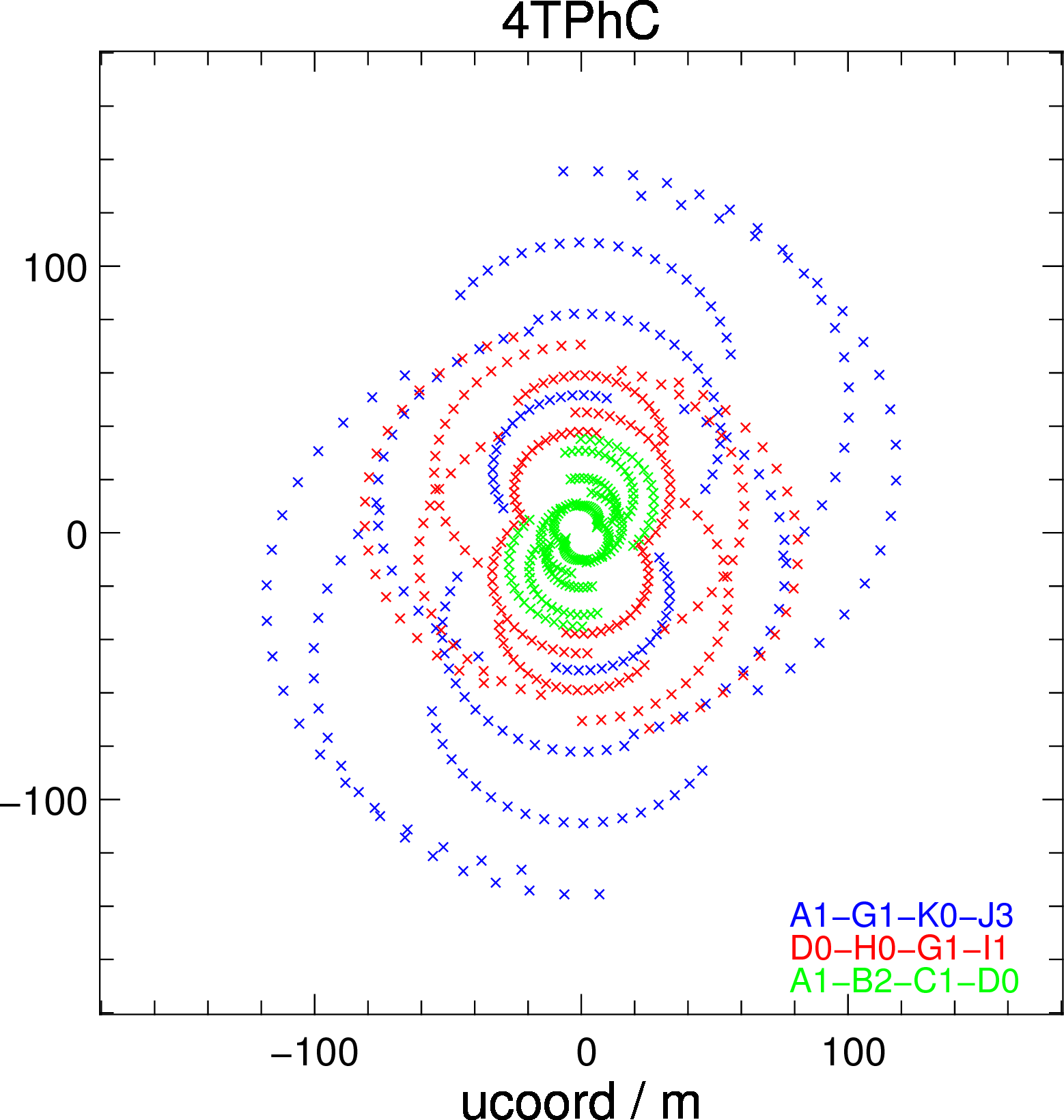}~
			\includegraphics[height=5.5cm]{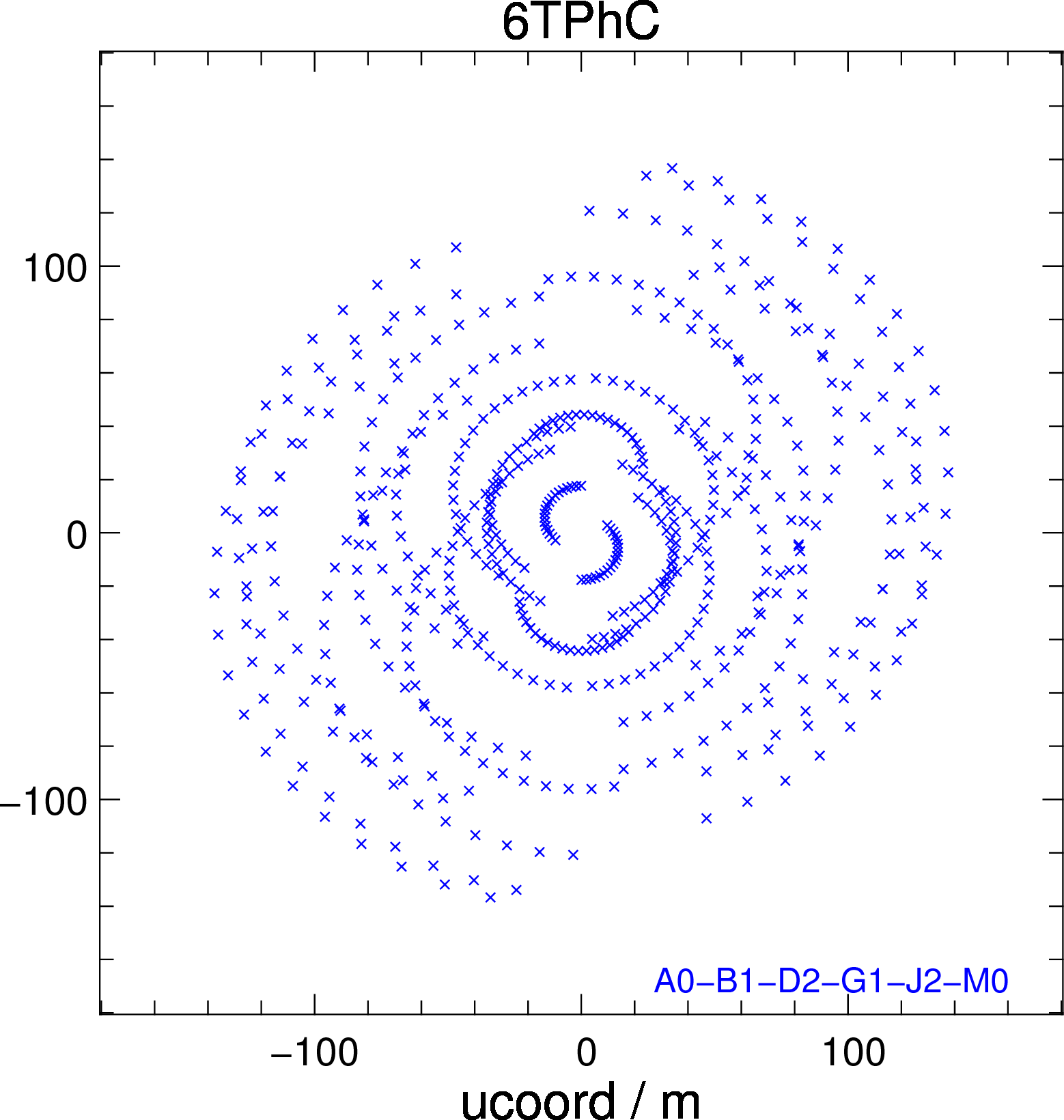}
			\caption{%
				\label{fig:uv-coverage}%
				$UV$-coverages of the observational configurations used in the study. %
				PhR stands for phase referencing and PhC for phase closure. %
				The observing nights are fixed for each column and are as follows: six nights left column, three nights central column and one night right column. %
				The stations used in each configuration are indicated.%
			}
			\end{center}
		\end{figure*}

		We used realistic $uv$-coverages for the VLTI station positions.\footnote{Available at \url{https://www.eso.org/observing/etc/doc/viscalc/vltistations.html}.}
		Six observational configurations are considered, corresponding to one, three and six nights of observation, and to phase referencing (PhR) and phase closure (PhC) data.
		The station configurations are inspired in previous imaging studies \citep{Filho2008a,Filho2008b}, and are representative of several instruments: \textit{PRIMA} (2TPhR; \citealt{Delplancke2008}), \textit{AMBER} (3TPhC; \citealt{Petrov2007}), \textit{GRAVITY} \citep[3T-4TPhR;][]{Eisenhauer2011}, \textit{PIONIER} (4TPhC; \citealt{LeBouquin2011,Eisenhauer2011}), and \textit{VSI} \citep[6TPhC;][]{Malbet2006}.
		To compute the $uv$-tracks, which depend on the object position, observatory location, station positions and hour-angle of the observations \citep{Thompson1986}, the following assumptions were made: (\textit{i}) object declination of $-60^\circ$, (\textit{ii}) a full $uv$-track corresponding to 19 instantaneous and evenly sampled data points, during a 9\,h transit, and (\textit{iii}) fixed station configurations during each night.
		The corresponding $uv$-coverages are presented in Fig.~\ref{fig:uv-coverage}.

		\subsection{Noise model}
		\label{sec:noise_model}
		The observables used in this study are the visibility amplitude $V$, the baseline visibility phase $\phi$, the squared visibility $V^2$, the bi-spectrum $\mathcal{B}$, and the closure phase $\phi_\MR{c}$.
		A synthetic observable $o_\MR{s}$ is generated by
		\begin{displaymath}
		\label{eq:synthetic_observable}
			o_\MR{s} \sim \mathcal{N}(\MR{E}\{o\},\MR{Var}\{o\}) ,
		\end{displaymath}
		where the expected value of the observable ($\MR{E}\{o\}$) is computed by interpolating the reference image at the angular frequencies of the observations,\footnote{The observing wavelength is taken at the centre of the \textit{K} band: $2.179\,\micron$.} using the \mira package.\footnote{Available for download at \url{http://cral.univ-lyon1.fr/labo/perso/eric.thiebaut/?Software/MiRA}.%
		} %
		We adopted the \textit{Simple Noise Model} \citep{Gomes2016b}, which is Gaussian and described by one free parameter, the signal-to-noise ratio ($\MR{SNR}$).
		It is assumed to be $\MR{SNR}= 20$, a value typical of good quality interferometric observations.
		The variance of the noise for the $n^\MR{th}$ visibility amplitude is defined as
		\begin{equation}
		\label{eq:sigma_V}
			\MR{Var}\{V_n\}= \left( \frac{\langle V \rangle}{\MR{SNR}} \right)^2 ,
		\end{equation}
		where $\langle V \rangle$ is the average of all visibility amplitudes for a given $uv$-coverage (\cf Table~\ref{tab:avg_visibilities}).
		
		In order to derive the noise for the baseline phase, we assume that the complex visibility has independent real and imaginary parts, with the same Gaussian noise \citep[Goodman approximation,][]{Goodman1985}.
		The variance of the noise for the $n^\MR{th}$ baseline phase becomes
		\begin{equation}
		\label{eq:sigma_phi}
			\MR{Var}\{\phi_n\}= \frac{\MR{Var}\{V_n\}}{V_n^2} .
		\end{equation}

		The noise for the remaining observables can be determined by error propagation. 
		
		The simple noise model is in contrast with the one used by \cite*{Renard2011}, since it initially sets the noise in the visibility amplitude instead of the phase, making the noise in the phase increase with decreasing visibility amplitude.
		It also qualitatively agrees with \cite{Tatulli2005}, where the visibility $\MR{SNR}$ ratio increases with the visibility amplitude.
		\begin{table*}
			\centering
			\begin{scriptsize}
				\arrs{1.3}
				\caption{%
					\label{tab:avg_visibilities}%
					Mean values of the distribution of the visibility amplitudes for the objects in each $uv$-configuration. %
					The errors correspond to the standard deviation.%
				}
				\begin{tabular}{c cc c cc c cc}
					\hline
					Object	& {2TPhR} & {3TPhC} & & {3TPhR} & {4TPhC} & & {4TPhR} & {6TPhC}\\
					\hline
					Stellar cluster
					& $0.6\pm0.2$ & $0.6\pm0.2$ &	& $0.6\pm0.2$ & $0.6\pm0.2$ & & $0.6\pm0.2$ & $0.6\pm0.2$ \\
					YSO
					& $0.4\pm0.2$ & $0.3\pm0.2$ & & $0.5\pm0.3$ & $0.5\pm0.3$ & & $0.4\pm0.2$ & $0.3\pm0.2$ \\
					Stellar photosphere
					& $0.3\pm0.2$ & $0.2\pm0.2$ & & $0.4\pm0.3$ & $0.4\pm0.3$ & & $0.3\pm0.2$ & $0.2\pm0.2$ \\
					\hline
				\end{tabular}
			\end{scriptsize}
		\end{table*}

		\subsection{Image reconstruction with \mira}
		\label{sec:mira_image_reconstruction}
		The generated noisy data are saved in an \softw{OIFITS} file \citep{Pauls2005} and used as input for the \mira image reconstruction software, assuming monochromatic data.
		As the goal of the study is to find the best metric for image reconstruction, the actual algorithm is not relevant, as long as it remains the same for all metrics.
		The \mira software and its principles are described in detail by \cite{Thiebaut2008,Thiebaut2013}.
		To summarize, \mira searches for the image $x^+$ which minimizes the two-term penalty criterion:
		\begin{equation}
		\label{eq:mira_image}
			x^+= \argmin_{x} \Bigl\{ f(x) = f_\MR{data}(x|d) + \mu \, f_\MR{prior}(x) \Bigr\} .
		\end{equation}
		The term $f_\MR{data}(x|d)$, usually known as the \textit{likelihood term}, measures the discrepancy between the actual data $d$ (\eg squared visibilities $V^2$, visibility amplitudes $V$, baseline phases $\phi$, and closure phases $\phi_\MR{c}$) and their model, given the image, $x$.
		The term $f_\MR{prior}(x)$, commonly designated as the \textit{regularization term}, is a penalty which enforces additional priors, and it is required to avoid artefacts.
		It is needed because the data alone cannot unambiguously yield a unique image.
		The so-called \textit{level of regularization} or \textit{hyper-parameter} $\mu> 0$ is adjusted to set the relative weight of the priors.
		In addition to minimizing the cost $f(x)$, the sought image $x^+$ is strictly constrained to be non-negative and normalized (the sum of the pixels being equal to 1).
		
		For the regularization term, we chose a relaxed version of the total variation criterion \citep*{Rudin1992,Strong2003}, which enforces edge-preserving smoothness \citep*{Charbonnier1997}, and that was found by \cite{Renard2011} to be the most effective for a large variety of astronomical objects:
		\begin{equation}
		\label{eq:total_variation}
			f_\MR{prior}(x)= \sum_{i,j}\sqrt{(x_{i+1,j} - x_{i,j})^2 + (x_{i,j+1} - x_{i,j})^2 + \varepsilon^2},
		\end{equation}
		with $x$ the image and $i, j$ the pixel indexes ($\varepsilon> 0$ is a small value to have a differentiable prior term).

			\subsubsection{Practical implementation}
			\label{sec:practical_implementation}
			Once the regularization is defined, \mira takes as input (\textit{i}) the data, (\textit{ii}) an optional initial estimate for the image -- assumed a square of $N\times N$ pixels -- (\textit{iii}) the pixel size $\delta\theta$, (\textit{iv}) the hyper-parameter $\mu$, and (\textit{v}) the maximum number of iterations.
			\mira stops once the convergence criterion is fulfilled or the maximum number of iterations is reached.
			It then outputs a reconstructed image.
			
			The image lateral size  is $\varOmega = N \, \delta\theta$.
			It provides a strict constraint which limits the support of the restored object and strongly impacts on the reconstruction process.
			As we want to have as few constraints as possible for the reconstruction, we chose an image size significantly larger than that of the object.
			In the present work, $\varOmega$ was set to be 40\,mas, roughly 2.5 times the object size.
			The pixel size should sample the maximum angular resolution in the Nyquist-Shannon sense, \ie $\delta\theta < \lambda/(2\,B_\MR{max})$, with $B_\MR{max}$ the maximum projected baseline length.
			However, it was found that to make image comparison of point-like structures reliable, a much smaller value had to be used: $\delta \theta \simeq \lambda/(12\,B_\MR{max})$.
			By combining the above constrains and taking into account that the maximum baseline of the configurations in Fig.~\ref{fig:uv-coverage} is $B_\MR{max} = 144\,\MR{m}$, we adopted $N = 160$ and $\delta\theta= 0.25\,\MR{mas}$.
			
			The only remaining parameters in the reconstruction are the (optional) initial  image estimate, the number of iterations, and the value of the hyper-parameter.
			Their joint management is described in the following subsection.

			\subsubsection{Tuning the hyper-parameter \texorpdfstring{$\mu$}{u}}
			\label{sec:tunning_mu}
			For the phase referencing reconstructions, \mira is called without an initial image estimate, which amounts to starting with a random guessing image whose pixels are drawn following an independent uniform law.
			For the phase closure restorations, the initial image was a quick reconstruction from the corresponding phase referencing observation\footnote{2TPhR for 3TPhC, 3TPhR for 4TPhC, and 4TPhR for 6TPhC.} with a large value of $\mu$.
			Because of the strong level of regularization, this image is a highly blurred version of the true image $z$.
			Other procedures could be devised to obtain the starting image for phase closure, such as a short image recover without any phase information, but this aspect is not important for the goal of this study, to wit, devise a method to assess the quality of final reconstructed images.
			The first restoration step (with or without initial guess) is performed for 300 iterations.
			
			The image reconstruction process then follows a cascade of calls\footnote{Each using 1000 iterations.} to \mira, where $\mu$ is reduced by a constant factor in each call.
			The intermediary restored image output in each step is used as the image estimate for the next call.
			The total number of calls in the cascade is five and seven, respectively, for PhR and PhC.\footnote{The two extra steps in the PhC case are necessary for better convergence and to properly centre the image in the FOV.}
			\mira normally achieves convergence before the maximum number of iterations is reached.
			In the PhC case, the initial image for the next \mira call was obtained by soft-thresholding the output of the previous call at 5\% of its maximum\footnote{$x_{k+1} = \max(0,\, x_k - 0.05\cdot\max(x_k))$, with $x_k$ the recovered image in step $k$.
				This approach was required because of the non-convex nature of PhC image reconstruction.
				The algorithm frequently converges to local minima.}.
			
			A limitation of the previous method is that convergence can be achieved for different values of $\mu$.
			Furthermore, no objective criterion for setting $\mu$ is available.
			In this work two approaches were followed to identify the best $\mu$.
			Initially, reconstructions were conducted for different values of $\mu$, spanning logarithmically from $10^4$ to $10^{-3}$.
			In the first approach, a human panel was asked to select the reconstructed image that most resembled the true image $z$, therefore determining the value of $\mu$. 
			In the second approach, the metrics selected in Section~\ref{sec:choice_candidates} were used.
			In our approach, the number of free parameters is kept to a minimum.
			In particular, we assume $\alpha=1$, $\beta=0$, a matching PSF $h=\delta$ (a Dirac function), and an effective PSF $h_\mathrm{ref}=G_{\sigma,t}$.
			The only free parameters are then the Gaussian $G$ standard deviation $\sigma$ and the translation $t$.
			The translation is only relevant for the closure phase case, where the object position cannot be determined from the data.
			Furthermore, the translation can be implemented in either $h$ or $h_\mathrm{ref}$.
			For simplicity it was implemented in $h_\mathrm{ref}$.
			The translation is not relevant for this work and will not be discussed further.
			The $\sigma$ is the only parameter of the metric expressing the effective resolution.
			Other functions (\eg Moffat functions) could be used, but as long as they reflect the shape of a PSF (characterized by a given width) the effect is not significant, because the metrics are summing over the convolved pixels of the images.
			By reducing the number of free parameters, this practical implementation has the further advantage of not defining a priori a given resolution for the reference image $y$, which could bias the results.
			Instead, it is a free parameter of the metric, that can be analysed later.
			The restored image $x$ is resampled to the grid of the reference image $y$.
			Then, each metric was evaluated in the 2D parameter space $(\mu,\sigma)$, with $\sigma$ spanning from 0 to 0.5\,mas.\footnote{For $\sigma=0$, the image is only shifted as expected from the analytic convolution.
				Because PhC does not keep the absolute position of the objects \citep{Monnier2007}, $h_\mathrm{ref}$ included a positional displacement $t = (t_1, t_2)$.
				This displacement was found by an iterative process that minimized the metric as a function of the displacement.}
			The minimum of the metric would then determine $\mu$.

	\section{Results and discussion}
	\label{sec:results}

		\subsection{Reconstructed images}
		\label{sec:reconstructed_images}
		\begin{figure*}
			\begin{center}
				\includegraphics[height=0.95\textheight]{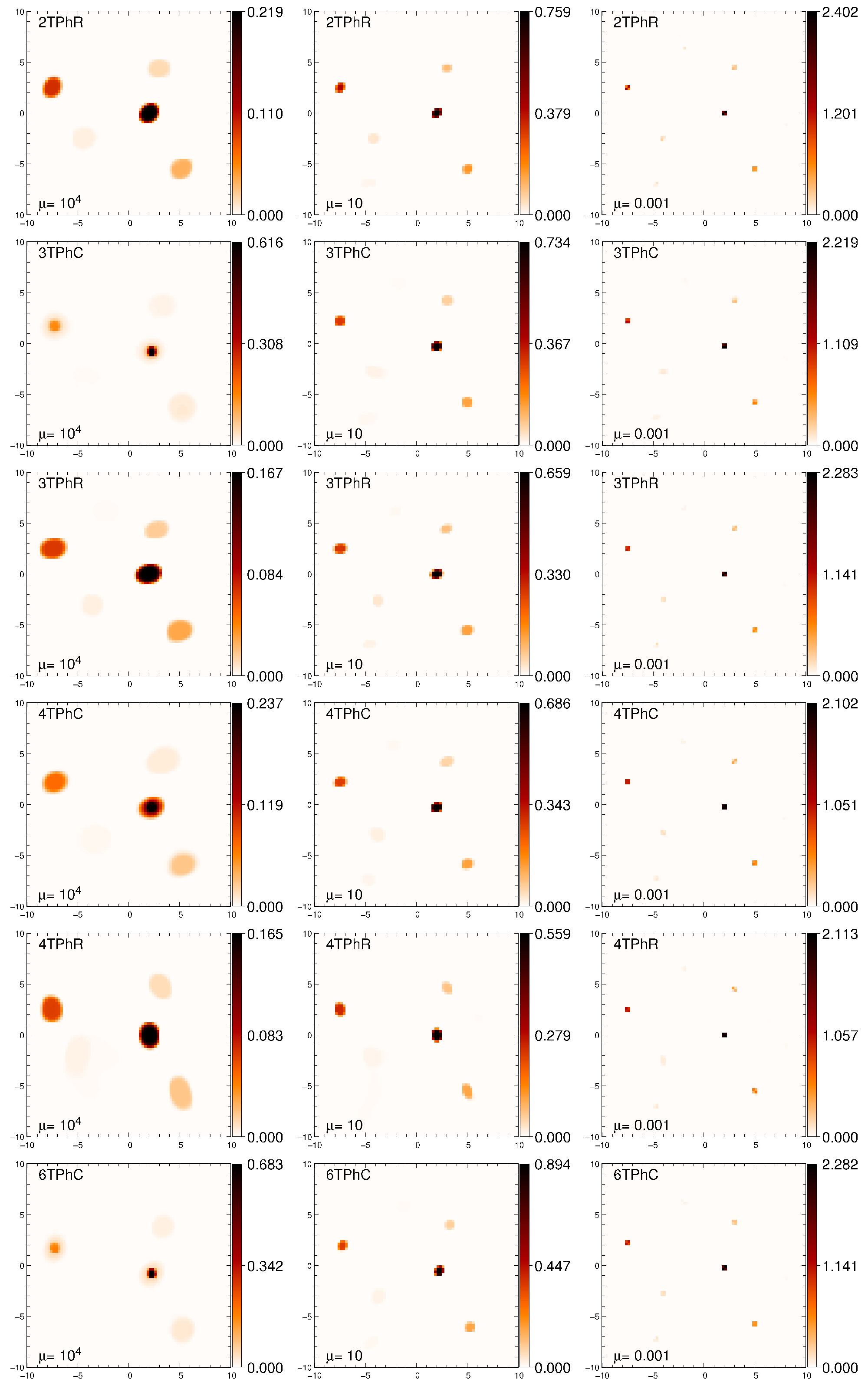}%
				\caption{%
					\label{fig:reconstructions_a-cluster} %
					Examples of image reconstructions for the stellar cluster. %
					Each column corresponds to a different level of regularization, and every row matches a different configuration of the synthetic observations. %
					The lateral image size is 20\,mas.%
				}
			\end{center}
		\end{figure*}
		\begin{figure*}
			\begin{center}
				\includegraphics[height=0.95\textheight]{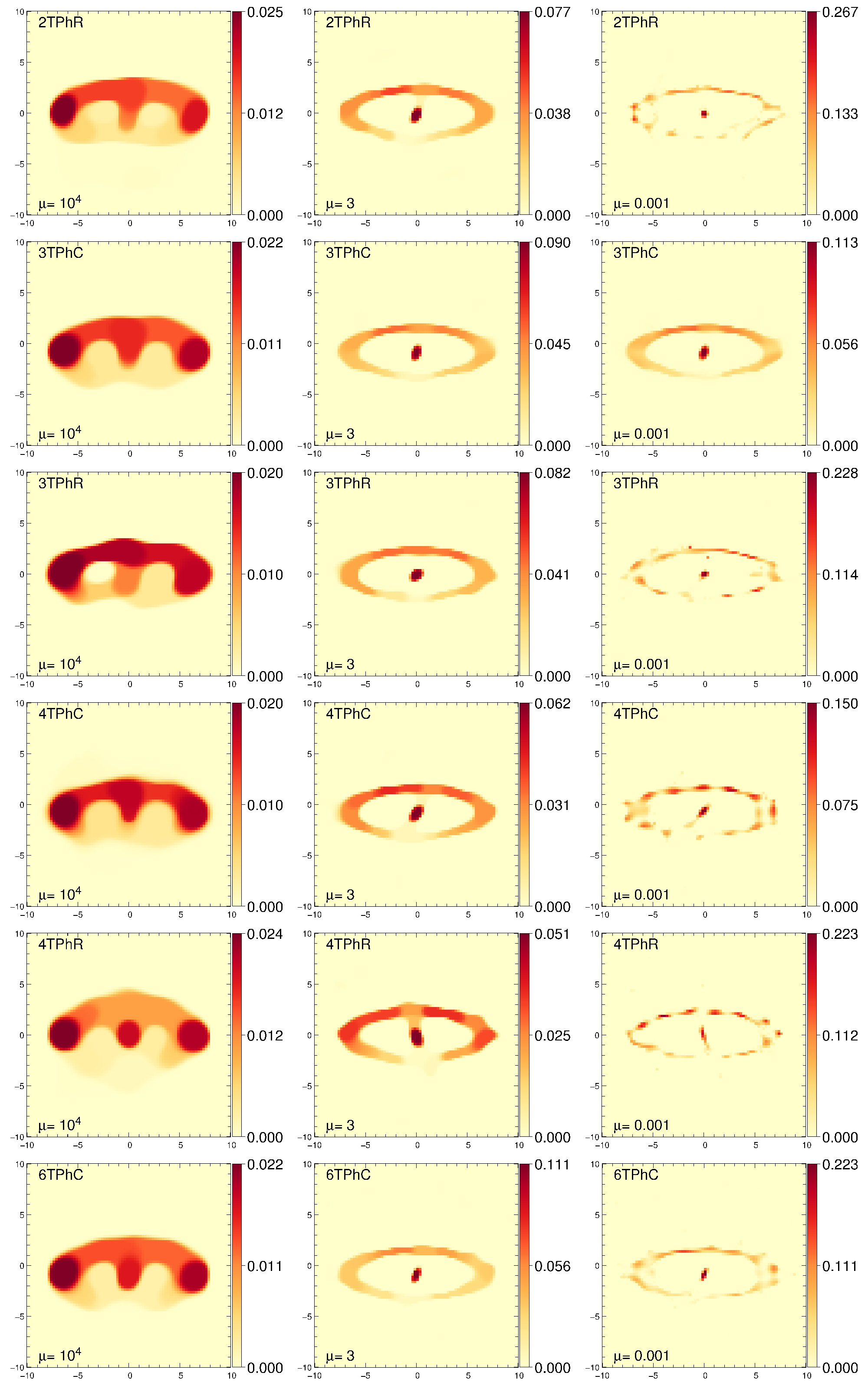}%
				\caption{%
					\label{fig:reconstructions_b-yso} %
					Same as in Fig.~\ref{fig:reconstructions_a-cluster}, but for the YSO.%
				}
			\end{center}
		\end{figure*}
		\begin{figure*}
			\begin{center}
				\includegraphics[height=0.95\textheight]{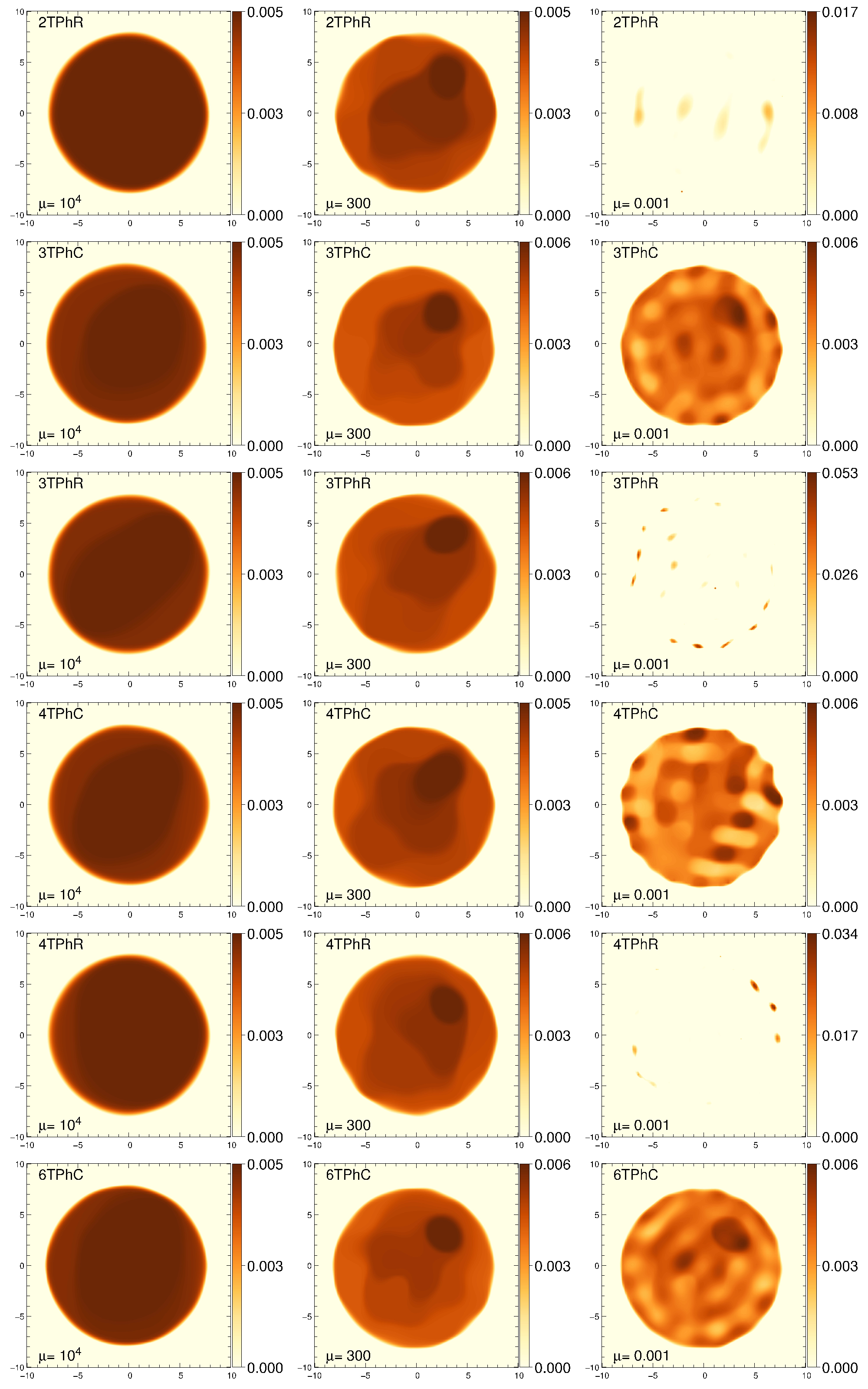}%
				\caption{%
					\label{fig:reconstructions_c-star} %
					Same as in Figs.~\ref{fig:reconstructions_a-cluster} and \ref{fig:reconstructions_b-yso}, but for the stellar photosphere.%
				}
			\end{center}
		\end{figure*}
		We produced 18 mock observations of the three reference images of Fig.~\ref{fig:reference} in all aforementioned array and phase scenarios.
		Images were restored from the corresponding interferometric data, stopping at 15 different levels of regularization, logarithmically ranging between $10^4$ and $10^{-3}$.
		The procedure was repeated twice, in order to create three sets of simulations and image reconstructions.
		Some examples of the 810 restored images are illustrated in Figs.~\ref{fig:reconstructions_a-cluster}--\ref{fig:reconstructions_c-star}.
		The full sets of recovered images are available at the JMMC website.\footnote{Available at \url{http://oidb.jmmc.fr/collection.html?id=gomes2016}.}
		
		Fig.~\ref{fig:reconstructions_a-cluster} corresponds to restored images of the stellar cluster, Fig.~\ref{fig:reconstructions_b-yso} to the YSO, and Fig.~\ref{fig:reconstructions_c-star} to the stellar photosphere.
		For the former, the first column lists images obtained when $\mu = 10^4$, the second column to $\mu = 10$, and the third column to $\mu = 10^{-3}$;
		for the YSO, the first column corresponds to $\mu = 10^4$, the second column to $\mu = 3$, and the last column to $\mu = 10^{-3}$;
		finally, for the stellar photosphere, $\mu = 10^4$ in the first column, $\mu = 300$ in the middle column, and $\mu = 10^{-3}$ in the last column.
		The rows are organized as follows: the phase cases alternate between PhR and PhC, and the number of telescopes increases from top to bottom -- two, three, four, and six telescopes (respectively 2T, 3T, 4T, and 6T) -- so as to get the scenarios 2TPhR, 3TPhC, 3TPhR, 4TPhC, 4TPhR, and 6TPhC.

		\subsection{Observational scenarios}
		\label{sec:observational_scenarios}
		The quality of the images changes according to the observational scenarios considered (2T, 3T, 4T and 6T, and PhR or PhC) and their respective $uv$-coverages.
		This is essentially related to the $uv$-coverage of the data and the amount of phase information.
		It is not the goal of the present study to compare phase referencing with phase closure (and the data presented do not allow us to draw conclusions), but to present a wide variety of situations in image reconstruction to successfully test merit functions.

		\subsection{Effect of the level of regularization on the image reconstruction}
		\label{sec:regularization}
		Concerning the reconstructions and levels of regularization (Figs.~\ref{fig:reconstructions_a-cluster}--\ref{fig:reconstructions_c-star}), it is noticeable that all restored images become sharper as the level of regularization is decreased, that is, as more weight is given to the data.
		However, below a certain level of $\mu$ -- which depends on the object and telescopes+phase configuration -- no visible effect on the shape and surface flux of the stellar cluster is seen, because the stars (point-like unresolved source objects) become confined to one pixel.
		This is not the case for objects with extended/resolved structures, such as the YSO and the stellar photosphere, where reducing the regularization below a certain level introduces reconstruction artefacts and noticeably degrades the quality of the image.
		For instance, in the YSO, for the highest tested level of regularization ($\mu = 10^4$) all images are blurred, with the central star attached to the disc.
		When $\mu = 3$, the disc is nicely restored in all configurations, with the central star separated from it.
		For $\mu = 10^{-3}$, only the 3TPhC configuration yields a well restored image.
		The configurations 2TPhR, 3TPhR, 4TPhC and 4TPhR exhibit disrupted discs, full of artefacts coming out of the reconstruction process, and the 6TPhC scenario produces an image where the disc, although intact, is very irregular.
		In the stellar photosphere, when $\mu = 10 ^4$, only the phase closure cases produce well enough restored images, with the most prominent spot visible.
		When $\mu = 300$, the 3TPhC and the 6TPhC cases yield images where the three spots are identifiable, but all other configurations produce discs full of restoration artefacts.
		For $\mu = 10^{-3}$, the 3TPhC and 6TPhC produce well enough restored images, with two and three spots identifiable respectively in the former and the latter configurations.
		In the remainder of the scenarios, the image is not properly restored -- the disc is not produced, and the algorithm gives rise solely to restoration artefacts distributed in a circular configuration.
		
		\subsection{Human determination of the hyper-parameter}
		\label{sec:human_mu}
		\begin{table*}
			\centering
			\begin{footnotesize}
				\arrs{1.3}
				\caption{%
					\label{tab:mus}%
					Value of the hyper-parameter $\mu$ obtained by the human panel. %
					The given values are the median of the values chosen by the experts, while the first and third quartiles are indicated between brackets.%
				}
				\begin{tabular}{l r@{\,}l r@{\,}l @{\hspace*{10mm}} r@{\,}l r@{\,}l @{\hspace*{10mm}} r@{\,}l r@{\,}l}
					\hline
					Object & \multicolumn{2}{c}{2TPhR} & \multicolumn{2}{c}{3TPhC} & \multicolumn{2}{c}{3TPhR} & \multicolumn{2}{c}{4TPhC} & \multicolumn{2}{c}{4TPhR} & \multicolumn{2}{c}{6TPhC}\\
					\hline
					Stellar cluster & $10$ & $\left({}^{30}_{\Z3}\right)$ & $30$ & $\left({}^{150}_{\Z\Z3}\right)$ & $3$ & $\left({}^{30}_{\Z3}\right)$ & $3$ & $\left({}^{30}_{\Z3}\right)$ & $10$ & $\left({}^{30}_{10}\right)$ & $10$ & $\left({}^{150}_{\Z10}\right)$\\
					YSO & $1$ & $\left({}^{1}_{1}\right)$ & $0.1$ & $\left({}^{1}_{0.001}\right)$ & $3$ & $\left({}^{10}_{\Z3}\right)$ & $3$ & $\left({}^{3}_{1}\right)$ & $3$ & $\left({}^{3}_{1}\right)$ & $3$ & $\left({}^{10}_{\Z3}\right)$\\
					Stellar photosphere & $300$ & $\left({}^{1000}_{\Z100}\right)$ & $300$ & $\left({}^{1000}_{\Z100}\right)$ & $300$ & $\left({}^{1000}_{\Z100}\right)$ & $1000$ & $\left({}^{1000}_{\Z100}\right)$ & $300$ & $\left({}^{300}_{300}\right)$ & $100$ & $\left({}^{300}_{\Z10}\right)$\\
					\hline
				\end{tabular}
			\end{footnotesize}
		\end{table*}

		Table~\ref{tab:mus} presents the average and standard deviation of the regularization hyper-parameter $\mu$ determined by the human panel, for each object and configuration.
		The value of $\mu$ for the stellar photosphere is much larger than for the stellar cluster, which in turn is larger than that for the YSO.
		For a given object, $\mu$ varies across configurations, without any specific pattern.

		\begin{table*}
			\centering
			\arrs{1.3}
			\caption{%
				\label{tab:sigmas_snr20}%
				Mean values of the $h_\mathrm{ref}$ $\sigma$ for the synthesized objects, observational scenarios and merit functions. %
				The numbers between parenthesis correspond to the standard error of the mean on the last digit.%
			}
			\begin{tabular}{lc cc c cc c cc}
				\hline
				& & \multicolumn{8}{c}{$\sigma$ / mas} \\
				\cline{3-10}
				Metric & & {2TPhR} & {3TPhC} & & {3TPhR} & {4TPhC} & & {4TPhR} & {6TPhC} \\
				\hline
				\multicolumn{10}{l}{\textit{Stellar cluster}}\\[1mm]
				ACC	 & & 0.14612(3) & 0.1484(3) & & 0.1481(2) & 0.1472(2) & & 0.1587(9) & 0.1481(1)\\
				L1N	 & & 0.14373(7) & 0.1483(4) & & 0.1464(4) & 0.1458(3) & & 0.1625(9) & 0.1498(3)\\
				L2N	 & & 0.14985(3) & 0.1522(3) & & 0.1518(2) & 0.1508(2) & & 0.1629(9) & 0.1520(1)\\
				IBC	 & & 0.15437(3) & 0.1560(3) & & 0.1560(2) & 0.1550(1) & & 0.1648(8) & 0.15547(9)\\
				\hline
				\multicolumn{10}{l}{\textit{YSO}}\\[1mm]
				ACC	 & & 0.281(2) & 0.273(4) & & 0.294(2) & 0.281(2) & & 0.320(2) & 0.263(2)\\
				L1N	 & & 0.204(2) & 0.191(5) & & 0.198(2) & 0.216(4) & & 0.259(4) & 0.207(2)\\
				L2N	 & & 0.306(3) & 0.298(4) & & 0.320(1) & 0.301(2) & & 0.347(2) & 0.282(2)\\
				IBC	 & & 0.343(4) & 0.333(4) & & 0.367(2) & 0.334(2) & & 0.384(2) & 0.305(3)\\
				\hline
				\multicolumn{10}{l}{\textit{Stellar photosphere}}\\[1mm]
				ACC	 & & 0.293(2) & 0.216(3) & & 0.270(2) & 0.255(2) & & 0.242(2) & 0.189(4)\\
				L1N	 & & 0.274(2) & 0.198(3) & & 0.239(2) & 0.232(2) & & 0.219(2) & 0.166(4)\\
				L2N	 & & 0.269(2) & 0.198(2) & & 0.245(2) & 0.233(2) & & 0.221(2) & 0.170(3)\\
				IBC	 & & 0.277(2) & 0.201(3) & & 0.251(2) & 0.239(3) & & 0.226(2) & 0.173(4)\\
				\hline
			\end{tabular}
		\end{table*}
		\begin{table*}
			\centering
			\arrs{1.3}
			\caption{%
				\label{tab:results-quality}%
				Mean values of the merit functions at the positions of $\mu$ determined by human selection (pink stars in Figs.~\ref{fig:merit_plots__cluster}--\ref{fig:merit_plots__star}). %
				The scores were obtained by computing the statistics for at least 12 realizations in each scenario. %
				The smaller the values, the better the agreement. %
				The numbers between parenthesis correspond to the standard error of the mean of the last digit.%
			}
			\begin{tabular}{lc cc c cc c cc}
				\hline
				& & {2TPhR} & {3TPhC} & & {3TPhR} & {4TPhC} & & {4TPhR} & {6TPhC} \\
				\hline
				\multicolumn{9}{l}{\textit{Stellar cluster}}\\[1mm]
				ACC	&	& 0.03760(9) & 0.0364(2) &	& 0.065(5) & 0.060(4) &	& 0.066(4) & 0.063(3) \\
				L1N	&	& 0.239(1) & 0.231(2) &	& 0.19(1) & 0.199(9) &	& 0.191(8) & 0.195(7) \\
				L2N	&	& $6.08(1)\times10^{-8}$ & $5.73(6)\times10^{-8}$ &	& $2.3(3)\times10^{-8}$ & $2.7(2)\times10^{-8}$ &	& $2.0(2)\times10^{-8}$ & $2.3(2)\times10^{-8}$ \\
				IBC	&	& $4.76(1)\times10^{-5}$ & $4.49(4)\times10^{-5}$ &	& $2.0(2)\times10^{-5}$ & $2.3(2)\times10^{-5}$ &	& $1.8(1)\times10^{-5}$ & $2.0(1)\times10^{-5}$ \\
				\hline
				\multicolumn{9}{l}{\textit{YSO}}\\[1mm]
				ACC	&	& 0.064(7) & 0.092(7) &	& 0.067(4) & 0.077(4) &	& 0.066(3) & 0.072(3) \\
				L1N	&	& 0.254(6) & 0.274(5) &	& 0.207(9) & 0.220(8) &	& 0.200(7) & 0.208(7) \\
				L2N	&	& $4.3(4)\times10^{-8}$ & $2.9(3)\times10^{-8}$ &	& $2.5(2)\times10^{-8}$ & $2.2(2)\times10^{-8}$ &	& $2.2(2)\times10^{-8}$ & $2.0(2)\times10^{-8}$ \\
				IBC	&	& $3.5(2)\times10^{-5}$ & $2.5(2)\times10^{-5}$ &	& $2.2(2)\times10^{-5}$ & $2.0(1)\times10^{-5}$ &	& $2.0(1)\times10^{-5}$ & $1.8(1)\times10^{-5}$ \\
				\hline
				\multicolumn{9}{l}{\textit{Stellar photosphere}}\\[1mm]
				ACC	&	& 0.083(6) & 0.068(5) &	& 0.074(4) & 0.068(4) &	& 0.070(3) & 0.066(3) \\
				L1N	&	& 0.24(1) & 0.19(1) &	& 0.208(8) & 0.187(8) &	& 0.201(7) & 0.187(7) \\
				L2N	&	& $2.5(3)\times10^{-8}$ & $1.9(3)\times10^{-8}$ &	& $2.0(2)\times10^{-8}$ & $1.8(2)\times10^{-8}$ &	& $2.0(2)\times10^{-8}$ & $1.8(1)\times10^{-8}$ \\
				IBC	&	& $2.2(2)\times10^{-5}$ & $1.7(2)\times10^{-5}$ &	& $1.8(1)\times10^{-5}$ & $1.6(1)\times10^{-5}$ &	& $1.8(1)\times10^{-5}$ & $1.6(1)\times10^{-5}$ \\
				\hline
			\end{tabular}
		\end{table*}

		The values of $\mu$ determined by human selection correspond to images that were fed to selected merit functions (see Section~\ref{sec:merit_functions}).
		The $h_\mathrm{ref}$ width is a remaining free parameter.
		We present in Table~\ref{tab:sigmas_snr20} the values of the Gaussian $\sigma$ that minimize the metric for the human determined $\mu$.
		These values were obtained by computing the statistics for 12 realizations in each object and observational scenario.
		The $\sigma$ values are of the order of 0.2\,mas, which corresponds to a full width at half-maximum of about 0.5\,mas.
		This should be compared to the angular resolution of the interferometer, which is around 3\,mas, and to the reference images pixel size of 0.25\,mas.
		Clearly the image reconstruction achieves a significant level of super-resolution, which is limited by the pixel size of the reconstructed images.
		This result might appear puzzling at first sight, but angular resolution is a sophisticated concept that cannot be fully enclosed in a simple Rayleigh-like criterion (\eg \citealt{denDekker1997}).
		Because we have prior information (enforced by the regularization and positivity of the solution), a reasonable SNR and relatively smooth objects, it is expected that the image reconstruction achieves significant super-resolution.
		
		In order to check the robustness of Figs.~\ref{fig:merit_plots__cluster} to \ref{fig:merit_plots__star} to different realizations of the data, we carried out 12 simulations of the 18 synthetic observations.
		The statistics of the minima for the human determined $\mu$ are presented in Table~\ref{tab:results-quality} (the errors in Table~\ref{tab:sigmas_snr20} were computed from this same data set).
		The standard error of the mean is very small, supporting the robustness of the results to the noise in the data set.

		\subsection{Benchmarking the metrics}
		\label{sec:benchmarking_metrics}
		\begin{figure*}
			\begin{center}
				\includegraphics[width=\textwidth]{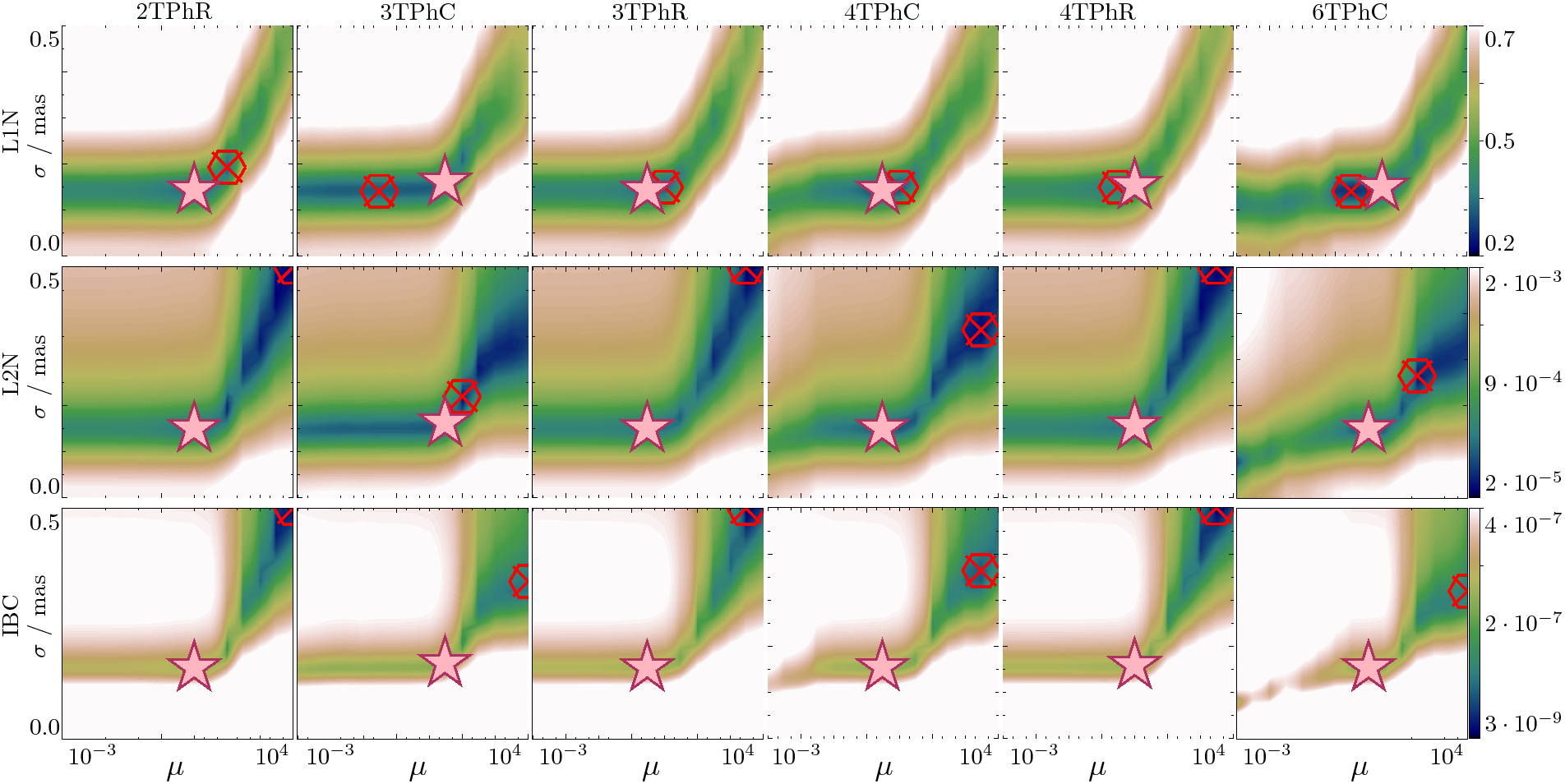}
				\caption{%
					\label{fig:merit_plots__cluster} %
					Average scores of the metrics L1N (\textit{top row}), L2N (\textit{central row}) and IBC (\textit{bottom row}) for three sets of simulated observations as function of the standard deviation $\sigma$ of $h_\mathrm{ref}$ and the level of regularization $\mu$. %
					The object is the stellar cluster of Fig.~\ref{fig:reference}. %
					From left to right, the panels are organized as follows: 2TPhR, 3TPhC, 3TPhR, 4TPhC, 4TPhR, and 6TPhC. %
					The red crossed circles correspond to global minima, while the pink stars are positioned at the human determined value of $\mu$ and the value of $\sigma$ that minimizes the merit function. %
					A logarithmic and a linear scale were respectively used for $\mu$ and $\sigma$.
				}
			\end{center}
		\end{figure*}
		\begin{figure*}
			\begin{center}
				\includegraphics[width=\textwidth]{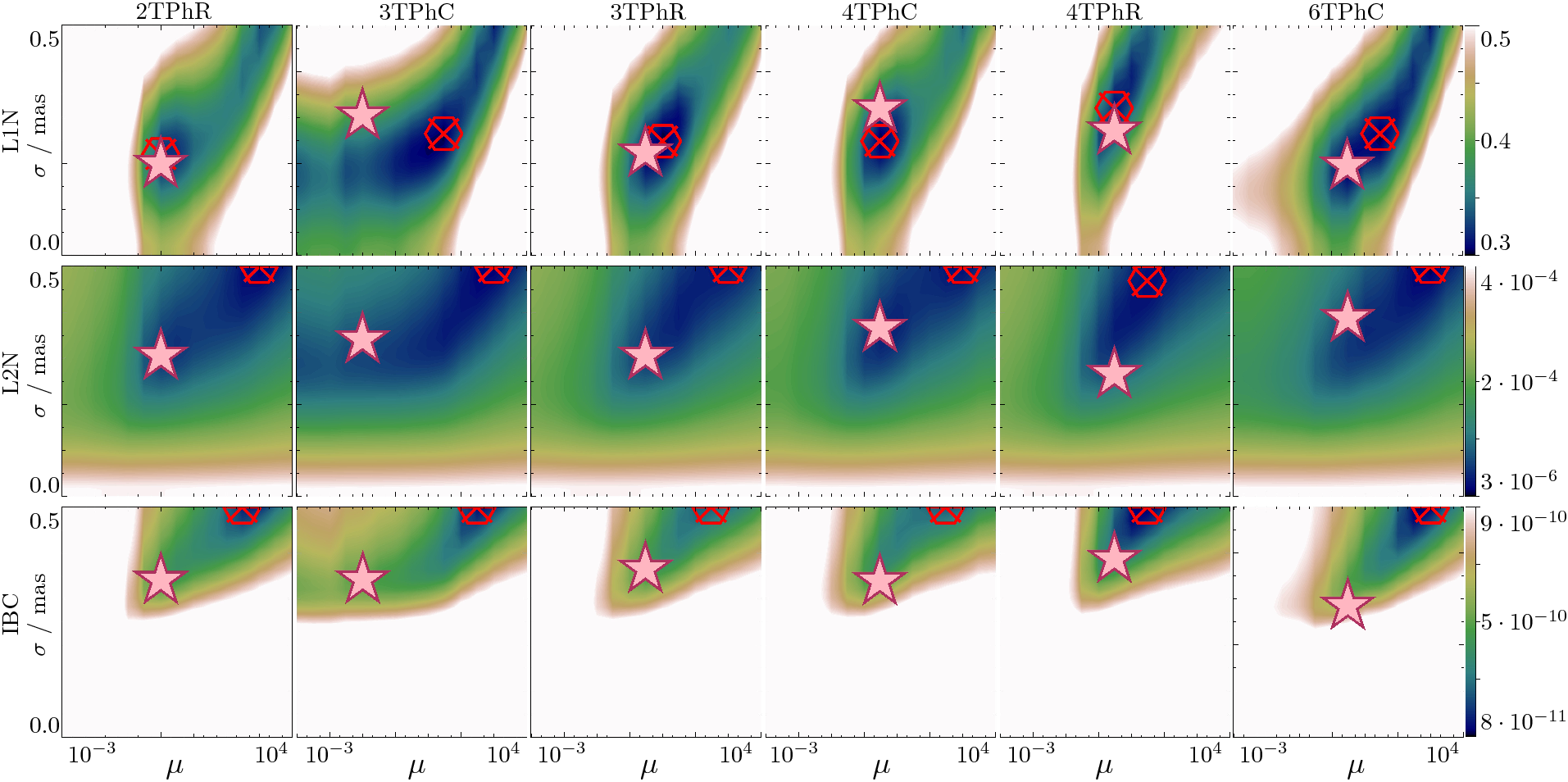}
				\caption{%
					\label{fig:merit_plots__yso} %
					Same as in Fig.~\ref{fig:merit_plots__cluster}, but for the YSO.%
				}
			\end{center}
		\end{figure*}
		\begin{figure*}
			\begin{center}
				\includegraphics[width=\textwidth]{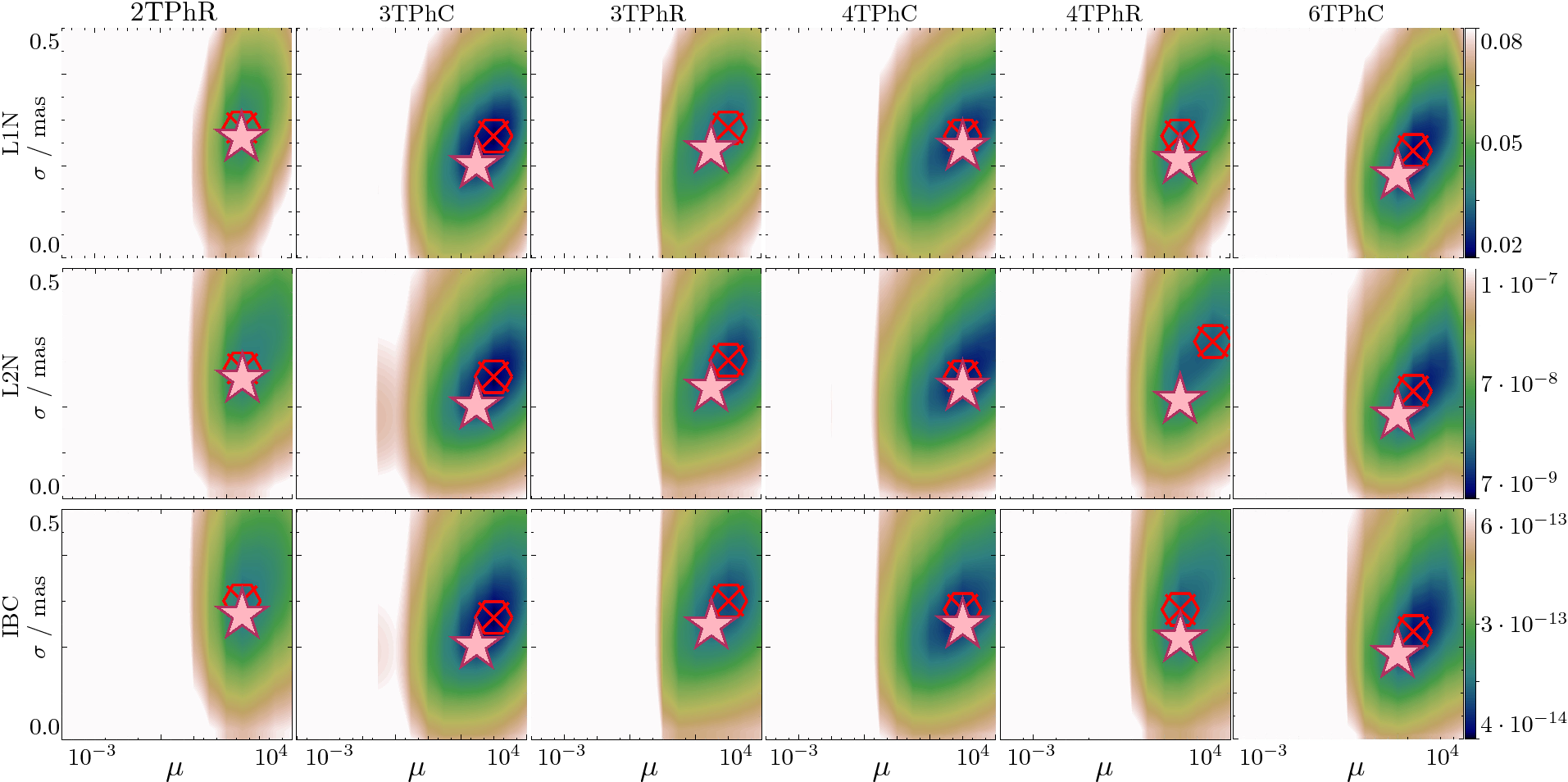}
				\caption{%
					\label{fig:merit_plots__star} %
					Same as in Figs.~\ref{fig:merit_plots__cluster} and \ref{fig:merit_plots__yso}, but for the stellar photosphere.%
				}
			\end{center}
		\end{figure*}

		As explained in Section~\ref{sec:mira_image_reconstruction}, a reconstructed image is a function of the final chosen $\mu$.
		Furthermore, the application of a given metric requires the convolution by $h_\mathrm{ref}$, whose width is characterized by $\sigma$.
		In this subsection we present and discuss the results for the behaviour of the merit functions.
		
		Table~\ref{tab:results-quality}, where $\mu$ is determined by human selection, provides an initial benchmark.
		The values of the quality functions show that IBC mimics the behaviour of L2N in most objects and configurations.
		On the one hand, this is explained by the quadratic nature of both metrics and, on the other hand, by the fact that the weighting function of IBC is the reference image itself, which makes the metric disregard pixels where the latter is zero.
		The failure of ACC in properly characterizing the quality of restored images in some scenarios is related to the fact that it applies a mask to the reference image before comparison, thus eliminating parts containing reconstruction artefacts that are important to determine the quality of the image.
		This however could be an interesting merit function when we are focused on certain parts of the image and want to eliminate others that we safely identify as artefacts of the reconstruction.
		For all objects and configurations, the L1N metric appears to properly characterize the quality of the restored images.
		
		We also conducted a systematic study of the metric behaviour as a function of $\mu$ and $\sigma$.
		We varied $\mu$ logarithmically between $10^4$ and $10^{-3}$, and $\sigma$ linearly between 0 and 0.5\,mas.
		The average values of the merit functions for three realizations of the simulated observations versus $\mu$ and $\sigma$ are plotted in Fig.~\ref{fig:merit_plots__cluster} (for the stellar cluster), Fig.~\ref{fig:merit_plots__yso} (for the YSO) and Fig.~\ref{fig:merit_plots__star} (for the stellar photosphere).
		The top, middle and bottom rows present the results for the quality functions L1N, L2N and IBC, respectively.
		The columns are organized as the rows of Figs.~\ref{fig:reconstructions_a-cluster}--\ref{fig:reconstructions_c-star}.
		The colour palette is inverted, such that the minima (darker colours) indicate a better agreement between the restored images and the references.
		All merit functions exhibit regions of minima, which is also verified in the ACC metric (not depicted).
		The red crossed circles point to the global minima of the panels.
		The pink stars are located at the position of the aforementioned values of $\mu$ determined by human selection.
		The position of the corresponding $\sigma$ was obtained by minimizing the merit function for the fixed $\mu$, using the \newuoa algorithm \citep{Powell2006}.
		
		The first result is that, generally, the merit functions are reasonably convex (\ie they depict regions with a clear minima).
		Overall, the effective resolution worsens with the hyper-parameter $\mu$, as expected (\ie the dark regions bend towards larger values of $\sigma$ and $\mu$).
		This is expected because increasing $\mu$ amounts to smooth the image.
		
		The shape of the minima regions of Figs.~\ref{fig:merit_plots__cluster} to \ref{fig:merit_plots__star} depends on the object.
		In the case of the stellar cluster (Fig.~\ref{fig:merit_plots__cluster}), the minima regions exhibit a horizontal branch up to a certain level of regularization.
		This is compatible with the aforementioned limiting value of regularization, below which restored images present no noticeable differences in quality and the (super-)resolution becomes limited by the size of the pixel.
		A single pixel encompasses the totality of the flux emanating from a restored unresolved star lying inside of it.
		The value of $\sigma \sim 0.15$\,mas indicated by the branch is compatible with the pixel size of 0.25\,mas.
		For sources with extended emission, the branch is not visible because the image degrades rapidly below a certain level of regularization (\cf Figs.~\ref{fig:reconstructions_b-yso} and \ref{fig:reconstructions_c-star} for some examples).
		Nevertheless, regions of minima are also evident, the position of which largely depends on the merit function.

			\subsubsection{L1N as the most robust metric}
			\label{sec:most_robust_metric}
			For L1N, the global minima typically lie well inside the limits defined by the plots.
			That is not the case for many L2N and IBC observations (especially for the cluster and YSO), suggesting that if the study was extended to larger values of $\sigma$ and $\mu$, the global minima would point to more blurred images.
			The minima valley oriented in the direction of increasing $\mu$ and $\sigma$ is less pronounced for L1N than for L2N and IBC.
			For L2N and IBC, this would indicate a better agreement between the restored and the reference images in those extreme regions of the plots, where the restored images are more blurred.
			This clearly shows that these metrics are biased and are not robust to over-smoothing by large values of the $\mu$ hyper-parameter.
			They will consider that an image with lower `angular resolution' is a better image than one with higher `angular resolution'.
			These results support L1N as the most robust of the merit functions used for the variety of cases considered.
			
			The morphology of the object has some impact on the behaviour of the metrics.
			The quality of extended resolved objects can be more easily assessed than that of unresolved sources.
			When the emitting source combines both types of objects (resolved and unresolved), the studied merit functions seem to have a harder job to evaluate the quality of the restored images.
			The great imbalance in intensity between the central star and the surrounding disc might explain the differences in quality.

			\subsection{Automatic image quality assessment}
			\label{sec:automatic_quality_assessment}
			The distance between the pink stars (minima obtained from human selection) and the circled red crosses (global minima) in Figs.~\ref{fig:merit_plots__cluster} to \ref{fig:merit_plots__star} indicates how well a given merit function translates the human perception of a ``good'' restored image.
			In this regard, L1N is clearly the best of all studied metrics, as it is the only one where both beacons lie close together for the typology of objects and observing configurations.
			
			This is not as well verified with the other metrics, being IBC the less robust of the tested merit functions.
			In the case of the stellar photosphere (Fig.~\ref{fig:merit_plots__star}), all metrics behave similarly.
			
			Since we are truncating the intervals of $\sigma$ and $\mu$, those distances most probably would increase in the cases where the global minima lie at extreme points of the plots.
			
			These results open the possibility of automatic image quality assessment, thus removing human intervention in the process.

	\section{Conclusions and future developments}
	\label{sec:conclusions}
	This article addresses the question: what is the best metric to assess the quality of a reconstructed image?
	
	Several merit functions are considered in the realistic context of the VLTI and using the \mira image reconstruction software. 
	
	A semi-automatic pipeline is developed to reconstruct images, with the only human intervention being the determination of the final value of the hyper-parameter $\mu$.
	It is found that the image reconstruction process outputs images with an effective angular resolution, characterized by a Gaussian, whose standard deviation $\sigma$ is significantly smaller than an equivalent Rayleigh-like criterion, based on the maximum baseline.
	Hence, a certain amount of super-resolution is achievable thanks to
the constraints imposed by a regularized image reconstruction algorithm.
	
	In order to cope with the mismatch between the effective resolution of the restored image and that of the simulated object, we advocate that convolution by an effective PSF is mandatory for proper image quality assessment.
	This effective PSF can be further used to compensate for image shift, which is unavoidable when image reconstruction is performed from power-spectrum and phase closure data.
	
	Of all the merit functions considered, the $\ell_1$-norm is the most robust.
	The commonly used Interferometric Imaging Beauty Contest quadratic metric is biased, considering as best images those with higher smoothing (or hyper-parameter $\mu$), and not fully exploiting the effective angular resolution of the data and image reconstruction process.
	
	By minimizing the $\ell_1$-norm over the $\mu$ and $\sigma$ parameter space, it is possible to implement automated image quality assessment.
	
	Based on this work, several developments are foreseen, the most obvious of which being algorithm comparison with the $\ell_1$-norm and proper convolution. The most ambitious is automated image reconstruction.
	To achieve this goal, two aspects must be addressed: (\textit{i}) the determination of an initial image for the reconstruction algorithm (for phase closure only), and (\textit{ii}) the determination of the final $\mu$ in the reconstruction.
	The second aspect is clearly the most difficult.
	It opens the requirements for image reconstruction algorithms to output tables of images for different levels of regularization, allowing the end-user to determine the final values of $\mu$.
	
	An important aspect is to identify the situations where phase referencing or phase closure are the best options for imaging.
	This choice is now possible with the \textit{GRAVITY} and \textit{PIONIER} instruments.
	Its study requires the inclusion of other ingredients not addressed in the present article, such as (\textit{i}) compatible $uv$-coverages, (\textit{ii}) noise models taking into account photon and detector statistics (\eg \citealt{Tatulli2005}) and/or light splitting between telescopes (\eg \citealt{Gordon2012}), and (\textit{iii}) a span of SNRs.

	\section*{Acknowledgements}
	\label{sec:acknowledgements}
	The research leading to these results has received funding from the European Community's Seventh Framework Programme, under Grant Agreements 226604 and 312430 (OPTICON), as well as from Fundação para a Ciência e Tecnologia grants SRFH/BD/44282/2008, PTDC/CTE-AST/116561/2010, UID/FIS/00099/2013 and COMPETE FCOMP-01-0124-FEDER-019965.
	This research has made use of the Jean-Marie Mariotti Center \softw{Aspro} service \footnote{Available at \url{http://www.jmmc.fr/aspro}}.
	All the analysis was done with \softw{YORICK}, a free interactive data processing language written by David Munro\footnote{Available at \url{http://yorick.sourceforge.net/}}.
	
	The authors would like to thank P.\ Andrade, N.\ Anugu, T.\ Armstrong, J.\ Ascenso, P.\ Berlioz-Arthaud, W.-J.\ de Wit, G.\ Duvert, S.\ Ertel, D.\ Faes, H.\ Hummel, P.\ Kervella, J.\ Kluska, Y.\ Kok, M.\ Langlois, J.\ Léger, M.\ Loupias, D.\ Mourard, M.\ Ozon, E.\ Pinho, N.\ Scott, M.\ Silva, F.\ Soulez, M.\ Tallon, I.\ Tallon-Bosc, and N.\ Verrier, for their valuable contribution in the poll aiming at determining the best levels of regularization for image reconstruction.
	
	The authors would also like to thank the referee for his/her insightful comments.

%
\bsp

\label{lastpage}

\end{document}